\documentclass[aps,prb,twocolumn,superscriptaddress]{revtex4}
\usepackage{graphicx}
\usepackage{amssymb}
\begin{document}

%\preprint{}
\title{Electron doping evolution of the magnetic excitations in NaFe$_{1-x}$Co$_{x}$As}

\author{Scott V. Carr}
\affiliation{Department of Physics and Astronomy, Rice University, Houston, Texas 77005, USA}
\author{Chenglin Zhang}
\affiliation{Department of Physics and Astronomy, Rice University, Houston, Texas 77005, USA}
\author{Yu Song}
\affiliation{Department of Physics and Astronomy, Rice University, Houston, Texas 77005, USA}
\author{Guotai Tan}
\affiliation{Department of Physics, Beijing Normal University, Beijing 100875, China}
\author{Yu Li}
\affiliation{Department of Physics and Astronomy, Rice University, Houston, Texas 77005, USA}
\author{D. L. Abernathy}
\affiliation{Quantum Condensed Matter Division, Oak Ridge National Laboratory, Oak Ridge, Tennessee 37831, USA}
\author{M. B. Stone}
\affiliation{Quantum Condensed Matter Division, Oak Ridge National Laboratory, Oak Ridge, Tennessee 37831, USA}
\author{G. E. Granroth}
\affiliation{Neutron Data Analysis and Visualization Division, Oak Ridge National Laboratory, Oak Ridge National Laboratory, Oak Ridge, Tennessee 37831, USA}
\author{T. G. Perring}
\affiliation{ISIS Facility, Rutherford Appleton Laboratory, Chilton, Didcot, Oxfordshire OX11 0QX, UK}
\author{Pengcheng Dai}
\email{pdai@rice.edu}
\affiliation{Department of Physics and Astronomy, Rice University, Houston, Texas 77005, USA}
\affiliation{Department of Physics, Beijing Normal University, Beijing 100875, China}


\begin{abstract}

We use time-of-flight (ToF) inelastic neutron scattering (INS) spectroscopy to investigate the doping dependence of magnetic excitations across the phase diagram of NaFe$_{1-x}$Co$_x$As with $x=0, 0.0175, 0.0215, 0.05,$ and $0.11$. The effect of electron-doping by partially substituting Fe by Co is to form resonances that couple with superconductivity, broaden and suppress  low energy ($E\le 80$ meV) spin excitations compared with spin waves in undoped NaFeAs. However, high energy ($E> 80$ meV) spin excitations are weakly Co-doping dependent. Integration of the local spin dynamic susceptibility $\chi^{\prime\prime}(\omega)$ of NaFe$_{1-x}$Co$_x$As reveals a total fluctuating moment of 3.6 $\mu_B^2$/Fe and a small but systematic reduction with electron doping. The presence of a large spin gap in the Co-overdoped nonsuperconducting NaFe$_{0.89}$Co$_{0.11}$As suggests that Fermi surface nesting is responsible for low-energy spin excitations. These results parallel Ni-doping evolution of spin excitations in BaFe$_{2-x}$Ni$_x$As$_2$, confirming the notion that  low-energy spin excitations coupling with itinerant electrons are important for superconductivity, while weakly doping dependent high-energy spin excitations result from localized moments.
\end{abstract}

\maketitle

\section{Introduction}

A common thread in high-transition temperature (high-$T_c$) 
copper oxides \cite{PALee,Eschrig,tranquada} and 
iron pnictides [Fig. 1(a)] \cite{kamihara,Stewart} is their close proximity to a static antiferomagnetic (AF) 
ordered parent compound \cite{scalapino,dai,dai15,inosov}.  Since magnetism may be responsible for many of the anomalous transport properties and origin of high-$T_c$ superconductivity in these materials \cite{scalapino}, previous efforts focused on understanding the evolution of 
magnetism as superconductivity is induced by electron or hole-doping 
to their AF parent compounds \cite{tranquada,dai,dai15,inosov}.
In the case of copper oxides, spin excitations in hole-doped superconductors are marked by an hourglass-like dispersion \cite{tranquada} and a neutron spin resonance coupled with superconductivity \cite{Eschrig}.  For iron pnictide superconductors \cite{Stewart}, much work over the past several years has focused on understanding the hole-and electron- doping evolution of spin excitations in BaFe$_2$A$_2$ due to the available large single crystals of these materials suitable for inelastic neutron scattering (INS) experiments \cite{mdlumsden,sxchi,sli09,dkpratt09,adChristianson,dsinosov09,mywang10,clester10,jtpark10,hfli10,dsinosov11,mwang11,lharriger,kmatan,msliu12,hqluo12,gstucker12,mgkim13,mwang13,hqluo13,Ibuka,MGKim15}. 
In the undoped state, BaFe$_2$As$_2$ forms a collinear AF structure similar to those shown in Fig. 1(b) below $T_N\approx 140$ K, narrowly preceded by a tetragonal-to-orthorhombic structural phase also below $T_s\approx 140$ K ($T_N\leq T_s$) \cite{cruz,qhunag}. 
Because of the twinned domains, each orthorhombic and perpendicular to each other [Fig. 1(b)], low-energy spin waves in single crystal BaFe$_2$As$_2$ are centered around
both AF ordering wave vectors ${\bf Q_{AF}}=(\pm1,0)$ and $(0,\pm1)$, respectively, in reciprocal space [Fig. 1(d)].
INS measurements using neutron time-of-flight (ToF) chopper spectrometers have shown that spin waves
of BaFe$_2$As$_2$
 extend to about $\sim$300 meV with 
local dynamic susceptibility, defined as wave vector integrated spin dynamic susceptibility over the dashed diamond area in Fig. 1(d) \cite{dai15}, 
peaking around 200 meV \cite{lharriger}.  When Co and Ni are doped into  BaFe$_2$As$_2$, partially replacing Fe and contributing additional electrons to the FeAs layer,superconductivity is induced \cite{Lester09,Pratt09,Christianson09,Nandi10,hqluo,xylu13} and static order is gradually suppressed. Additionally, the low-energy ($E<100$ meV) spin excitations become broader than the spin waves in undoped BaFe$_2$As$_2$ and couple with superconductivity in the form of a neutron spin resonance similar to the superconducting copper oxides \cite{mdlumsden,sxchi,sli09,dkpratt09,adChristianson,dsinosov09,mywang10,clester10,jtpark10,hfli10,dsinosov11,mwang11}.
However, high-energy ($E>100$ meV) spin excitations remain weakly electron-doping dependent and 
are reminiscent of spin waves in the undoped BaFe$_2$As$_2$ \cite{msliu12,hqluo12,gstucker12,mgkim13,mwang13,hqluo13}.
In concert, these results suggest that low-energy spin excitations in electron-doped BaFe$_2$As$_2$ family of materials  
arise from itinerant electrons and Fermi surface nesting \cite{mazin2011n,Hirschfeld,Chubukov}, while high-energy spin excitations are related to local moments and are insensitive to changes in Fermi surfaces \cite{msliu12,mwang13,hqluo13,khaule08,qmsi08,cfang08,ckxu08}.

Although INS experiments on BaFe$_2$As$_2$ family of iron pnictides over the past several years have 
established the basic characteristics of the electron and hole-doping evolution of 
spin excitations and their coupling to superconductivity \cite{mdlumsden,sxchi,sli09,dkpratt09,adChristianson,dsinosov09,mywang10,clester10,jtpark10,hfli10,dsinosov11,mwang11,lharriger,kmatan,msliu12,hqluo12,gstucker12,mgkim13,mwang13,hqluo13,Ibuka,MGKim15}, 
it is equally important to determine if the features found in BaFe$_2$As$_2$ family of materials are universal for other iron pnictide superconductors.  For example, while the maximum $T_c$ ($\sim$ 20 K) for 
electron-doped NaFe$_{1-x}$Co$_x$As family of iron pnictides \cite{CWChu09,Parker10,AFWang12,GTTan13} is similar 
to that for Co/Ni-doped BaFe$_2$As$_2$ [Fig. 1(a)] \cite{Stewart}, it is unclear if the electron-doping evolution of spin excitations in 
NaFe$_{1-x}$Co$_x$As also behaves similarly to that of BaFe$_{2-x}$(Co,Ni)$_x$As$_2$ \cite{mdlumsden,sxchi,sli09,dkpratt09,adChristianson,dsinosov09,mywang10,clester10,jtpark10,hfli10,dsinosov11,mwang11,lharriger,kmatan,msliu12,hqluo12,gstucker12,mgkim13,mwang13,hqluo13,Ibuka}.
From INS experiments on spin waves in the undoped NaFeAs \cite{SLi09a}, we know that total magnetic bandwidth in NaFeAs is considerably smaller than that of BaFe$_2$As$_2$ \cite{CLZhang14}.
This is consistent with the density functional
theory (DFT) combined with dynamical mean field theory
(DMFT) calculation that the increased iron pnictogen 
height in NaFeAs increases the electron correlations (localizations) and narrows
spin-wave bandwidth compared with that of BaFe$_2$As$_2$ \cite{Kotliar06,Haule10,ZPYin14}.  If spin excitations are  
mediating the electron pairing for high-$T_c$ superconductivity, the superconducting condensation energy $U$
should be accounted for by the change in magnetic exchange energy
$\Delta E_{ex}(T)=2J[\left\langle {\bf S}_{i+x}\cdot{\bf S}_i\right\rangle_N-
\left\langle {\bf S}_{i+x}\cdot{\bf S}_i\right\rangle_S]$, where $J$ is the nearest neighbor magnetic exchange coupling and
$\left\langle {\bf S}_{i+x}\cdot{\bf S}_i\right\rangle$ is the magnetic scattering
in absolute units at the normal ($N$) and superconducting ($S$)
phases at zero temperature \cite{scalapino}, within an isotropic $t$-$J$ model \cite{Spalek}.
Since the effective magnetic exchange coupling constants of NaFeAs \cite{CLZhang14} are considerably smaller 
than those of BaFe$_2$As$_2$ \cite{lharriger}, it will be instructive to systematically map out the overall 
spin excitations spectra in NaFe$_{1-x}$Co$_x$As and compare the result with 
those of BaFe$_{2-x}$Ni$_x$As$_2$ family of materials \cite{msliu12,hqluo12,mwang13,hqluo13}.
In previous INS experiments on NaFe$_{1-x}$Co$_x$As probing 
low-energy spin excitations using triple-axis spectrometry, we find the presence of a single, sharp neutron spin resonance in a sample with nearly optimal Co-doping, similar to the resonance in electron-doped BaFe$_2$As$_2$ \cite{CLZhang13a}, while an underdoped sample with coexisting superconductivity and AF order exhibits a double resonance \cite{CLZhang13b,CLZhang14a}.
To illuminate the rest of the story, the Co-doping evolution of high-energy spin excitations in superconducting and non-superconducting NaFe$_{1-x}$Co$_x$As needs to be established.

\begin{figure}[t!]
	\centering
	\includegraphics[scale=1]{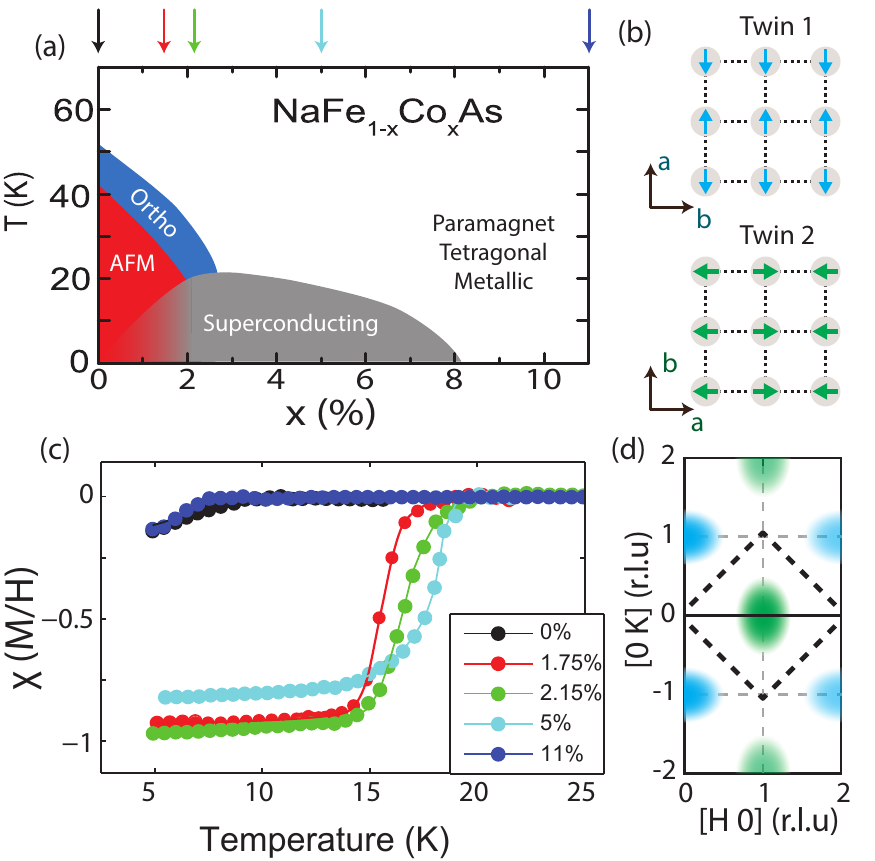}
	\caption{
		(Color online) (a) Schematic phase diagram of NaFeCoAs from thermodynamic measurements\cite{GTTan13}. Colored arrows above figure indicate doping values used in this paper. (b) In-plane magnetic order in twinned orthorhombic domains. (c) DC magnetic susceptibility $\chi$. (d) Neutron scattering schematic indicating intensity at [1,0] and [0,1] originate from different crystal domains.
	}
\end{figure}

In this article, we report ToF INS studies of temperature and doping dependence 
of spin excitations over the entire Brillouin Zone (B.Z.)
in NaFe$_{1-x}$Co$_x$As.  A schematic phase diagram of NaFe$_{1-x}$Co$_x$As is presented in Figure 1(a), where all high temperature compounds are paramagnetic metals with a tetragonal structure illustrated by the white region. The white-blue border indicates the tetragonal to orthorhombic structural transition and the red region depicts stripe AF order in the orthorhombic compound. Superconductivity exists in the gray region, where the opaqueness illustrates the superconducting volume fraction. When fully opaque, compounds in the superconducting region are tetragonal and not magnetically ordered. We chose Co-doping concentrations of $x=0,0.0125, 0.0175,0.05$, and 0.11, as shown by the arrows in the 
 electronic phase diagram of NaFe$_{1-x}$Co$_x$As [Fig. 1(a)]  \cite{CWChu09,Parker10,AFWang12,GTTan13}. 
Since NaFeAs has similar orthorhombic AF ground state as BaFe$_2$As$_2$ [Fig. 1(b)] \cite{SLi09a}, AF Bragg peaks and spin excitations from twinned domains will appear at ${\bf Q_{AF}}=(\pm1,0)$ and $(0,\pm1)$ positions in reciprocal space [Fig. 1(d)].  Figure 1(c) shows temperature dependence of the magnetic susceptibility $\chi$. While the $x=0$ and 0.11 samples are not bulk superconductors (the slight drop in susceptibility is due to filamentary superconductivity) \cite{AFWang12,GTTan13}, the $x=0.0175$ ($T_c\approx 16$ K) and 0.0215 ($T_c\approx 18$ K) samples are in the underdoped and nearly optimally doped
regime, $x=0.05$ is Co-overdoped with $T_c\approx 20$ K [Fig. 1(c)]. $T_c$ is estimated by the onset of steepest descent of $\chi$.  This range of Co-doped NaFeAs cover the entire superconducting phase diagram of the system, from 
undoped NaFeAs to underdoped, near optimally doped, overdoped superconducting and nonsuperconducting NaFe$_{1-x}$Co$_x$As.
Compared with spin waves in undoped NaFeAs, we find that Co-doping in NaFeAs slightly elongates the low-energy spin excitations along the
transverse direction around the commensurate AF order wave vector. 
For superconducting samples, a neutron spin resonance forms below $T_c$ consistent with earlier work \cite{CLZhang13a,CLZhang13b,CLZhang14a}. 
For Co-overdoped nonsuperconducting NaFe$_{0.89}$Co$_{0.11}$As, a large spin gap forms in the low-temperature 
state very similar to Ni-overdoped
nonsuperconducting BaFe$_{1.7}$Ni$_{0.3}$As$_2$ \cite{mwang13}. 
By comparing ToF INS data in NaFe$_{1-x}$Co$_x$As
with $x=0,0.0125, 0.0175,0.05$, we establish Co-doping evolution of the
wave vector and energy dependence of the
spin excitations throughout the B.Z. and find high-energy ($E>80$ meV) spin excitations are weakly Co-doping dependent.  
Although NaFe$_{1-x}$Co$_x$As family of materials has stronger electron correlations and weaker magnetic exchange couplings compared with 
those of BaFe$_2$As$_2$ based superconductors, superconductivity-induced changes in spin excitations are still much larger than the superconducting condensation energy.  These results are similar to those of BaFe$_2$As$_2$ based materials, and are consistent with idea 
that magnetic excitations are important for superconductivity of iron pnictide superconductors.   

\begin{figure}[t!]
	\centering
	\includegraphics[scale=0.85]{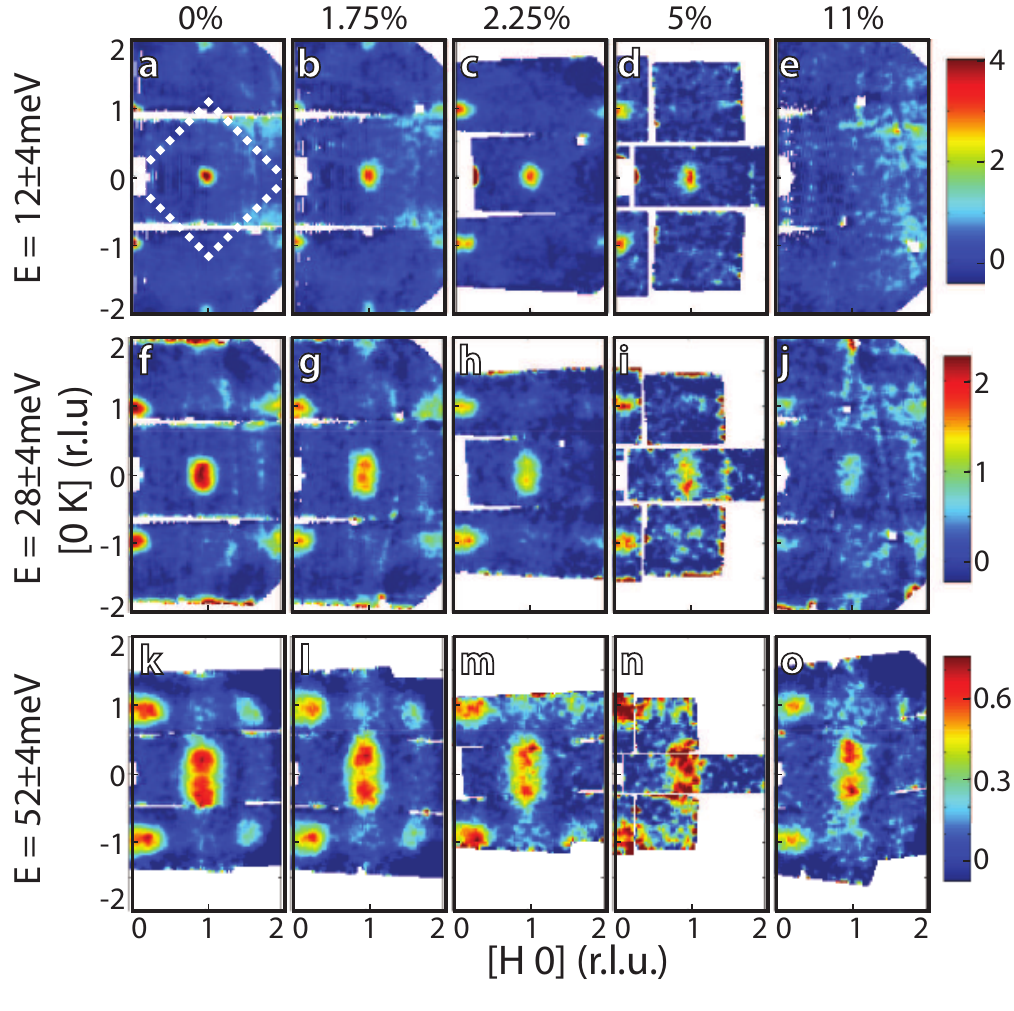}
	\caption{
		(Color online) Two dimensional slices from ToF INS of NaFe$_{1-x}$Co$_x$As measured with incident energy of 80meV at energy transfers of 12$\pm$4 meV (a-e), 28$\pm$4 meV (f-j), 52$\pm$4 meV (k-o) for doping values x=0, 0.0175, 0.0215, 0.05 and 0.11 respectively. White box in (a) indicates first magnetic B.Z.
	}
\end{figure}

\begin{figure}[t!]
	\centering
	\includegraphics[scale=1]{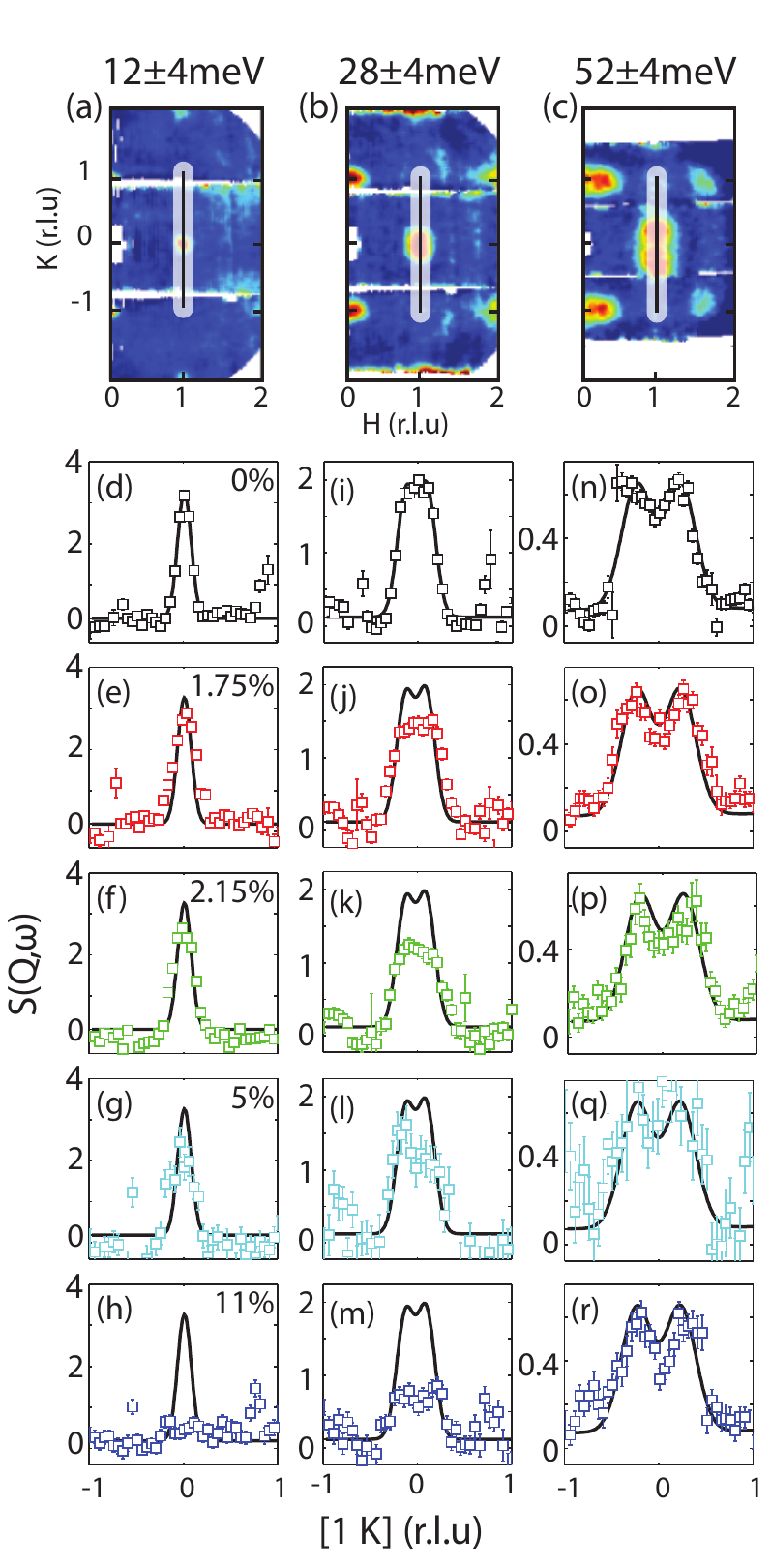}
	\caption{
		(Color online) (a-c) Illustrations of one dimensional transverse cuts from Fig. 2(a), 2(f), 2(k). Cuts measured with incident energy of 80meV at energy transfers of 12±4 meV (d)-(h), 28±4 meV (i)-(m), 52±4 meV (n)-(r) for doping values x=0, 0.0175, 0.0215, 0.05 and 0.11 respectively. All cuts integrated over $0.9<H<1.1$. Black curves are a Gaussian fit to the parent compound at identical energy transfer.
	}
\end{figure}

\section{Experimental Results}

Our ToF INS experiments were carried out at the
wide angular-range chopper spectrometer (ARCS) \cite{ARCS} and fine-resolution Fermi chopper spectrometer (SEQUOIA) \cite{SEQUOIA} 
at the Spallation Neutron Source (SNS), Oak Ridge National Laboratory (ORNL), and at MAPS chopper spectrometer 
at the Rutherford-Appleton Laboratory, UK. Large single crystals of NaFe$_{1-x}$Co$_x$As were grown by self-flux method \cite{CLZhang13a,CLZhang13b,CLZhang14a}.  Since these samples are highly air sensitive \cite{Spyrison12}, we have protected them with thin aluminum foil envelope coated with hydrogen-free 
amorphous fluoropolymer CYTOP. The mass of the CYTOP was negligible compared to the sample mass and no scattering features from the glue were observed for the energy range probed ($E>10$ meV).
To compare with spin wave results in undoped NaFeAs \cite{CLZhang14} and those of BaFe$_2$As$_2$ \cite{lharriger},
we define the wave vector $\bf Q$ at ($q_x$, $q_y$, $q_z$) as $(H,K,L) = (q_xa/2\pi, q_yb/2\pi, q_zc/2\pi)$ reciprocal lattice units (r.l.u.), where $a \approx b \approx 5.56$ \AA, and $c = 6.95$ \AA. For the $x=0,0.0175,$ and 0.11 compounds, we used the ARCS spectrometer. 
The experiments on the $x=0.0215$ and 0.05 compounds were carried out on the SEQUOIA and MAPS spectrometers, respectively.
 Crystals were co-aligned using CG-1B, a cold neutron alignment station at High-Flux Isotope Reactor (HFIR) 
and affixed to 
aluminum plates with aluminum wire. Each sample array was co-aligned in the $[H,0,L]$ scattering plane with a mosaic of less than 3 degrees. For each experiment, sample arrays with a total mass of 18-g, 11-g, 19-g, 15-g, and 10-g for $x=0, 0.0175, 0.0215, 0.05,$ and 0.11 respectively were loaded into a closed-cycle helium displex with the incident beam parallel to the $c$-axis to 
display the $[H, K]$ scattering plane. All measurements were performed at the base temperature of 5 K unless otherwise noted.
Each sample was measured at different subsets of incident energies in the range $E_i =25, 35, 50, 80, 150, 250, 350,$ and 450 meV, with all compounds measured with $E_i = 80$ and 250 meV. A detailed list of incindent energies measured for each doping can be found in Table II located in the appendix. For direct comparison of spin wave intensities between samples, each spectrum was first normalized to absolute units (mbarn/sr/meV/f.u.) using a vanadium standard to account for sample mass, then to each other by relative phonon intensity to account for spectrometer differences and residual flux in the sample (see appendix). Finally, phonon self-normalization was used to confirm the magnitude of spin excitation intensities\cite{PhonNorm}. 

The neutron scattering function $S(Q,E)$ is related to the imaginary part of the spin 
dynamic susceptibility $\chi^{\prime\prime}(Q,E)$ by correcting for the Bose population factor
via $S(Q,E)= 1/(1-\exp(-E/(k_BT)))\chi^{\prime\prime}(Q,E)$, where $k_B$ is the
Boltzmann's constant.  We can then calculate the local dynamic susceptibility by using
$\chi^{\prime\prime}(E)=\int{\chi^{\prime\prime}({\bf
    Q},E)d{\bf Q}}/\int{d{\bf Q}}$ (in units of $\mu_B^2$/eV/f.u.) with the integration over the 
B.Z. noted by the white outlined region in Fig. 2(a),
where $\chi^{\prime\prime}({\bf Q},E)=(1/3) tr( \chi_{\alpha \beta}^{\prime\prime}({\bf Q},E))$ \cite{clester10,msliu12}.
All calculations require the background to be determined and subtracted before computation. A detailed description of background calculation and subtraction is outlined in the appendix.

\begin{figure}[t!]
\includegraphics[scale=0.85]{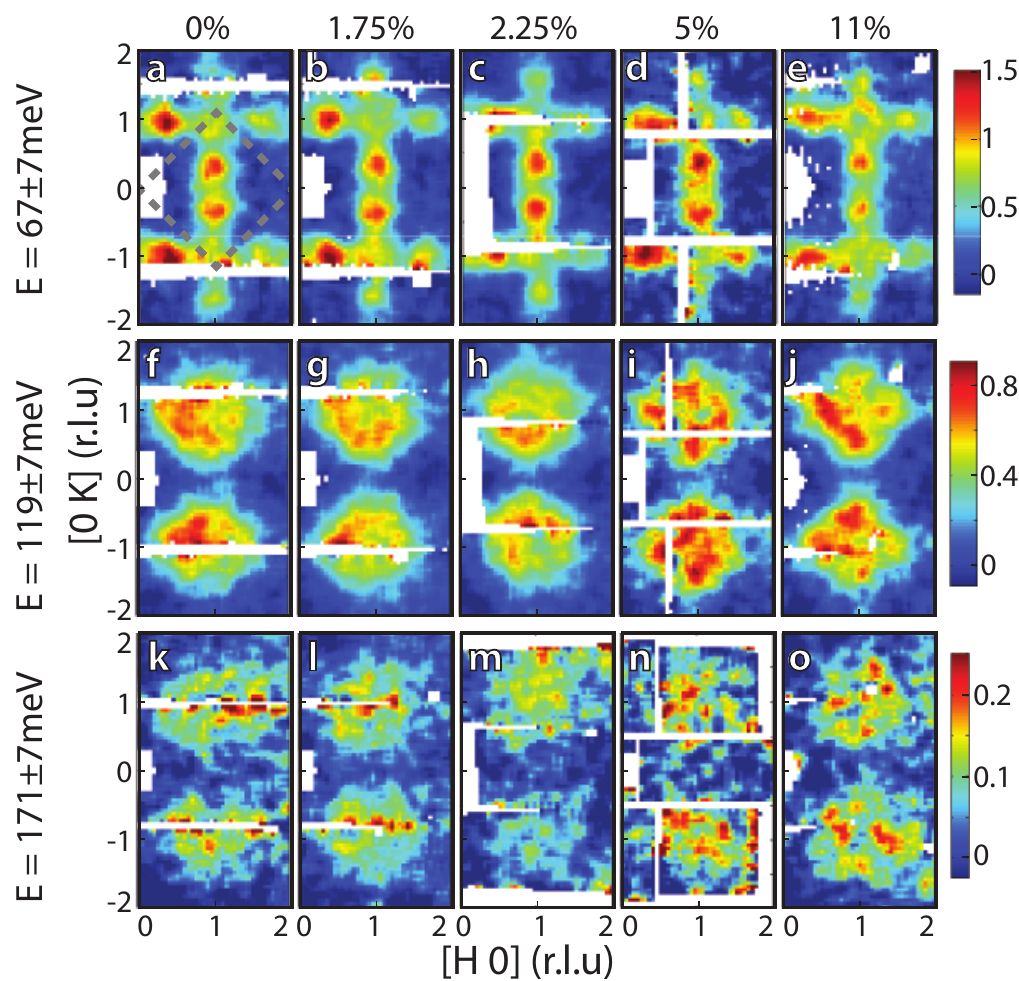}
\caption{
(Color online) 
Two dimensional slices from ToF INS of NaFe$_{1-x}$Co$_x$As measured with incident energy of 250meV at energy transfers of 67$\pm$13 meV (a-e), 119$\pm$13 meV (f-j), 171$\pm$13 meV (k-o) for doping values $x=0, 0.0175, 0.0215, 0.05$, and 0.11, respectively. 
}
\end{figure}

\begin{figure}[t!]
	\centering
	\includegraphics[scale=1]{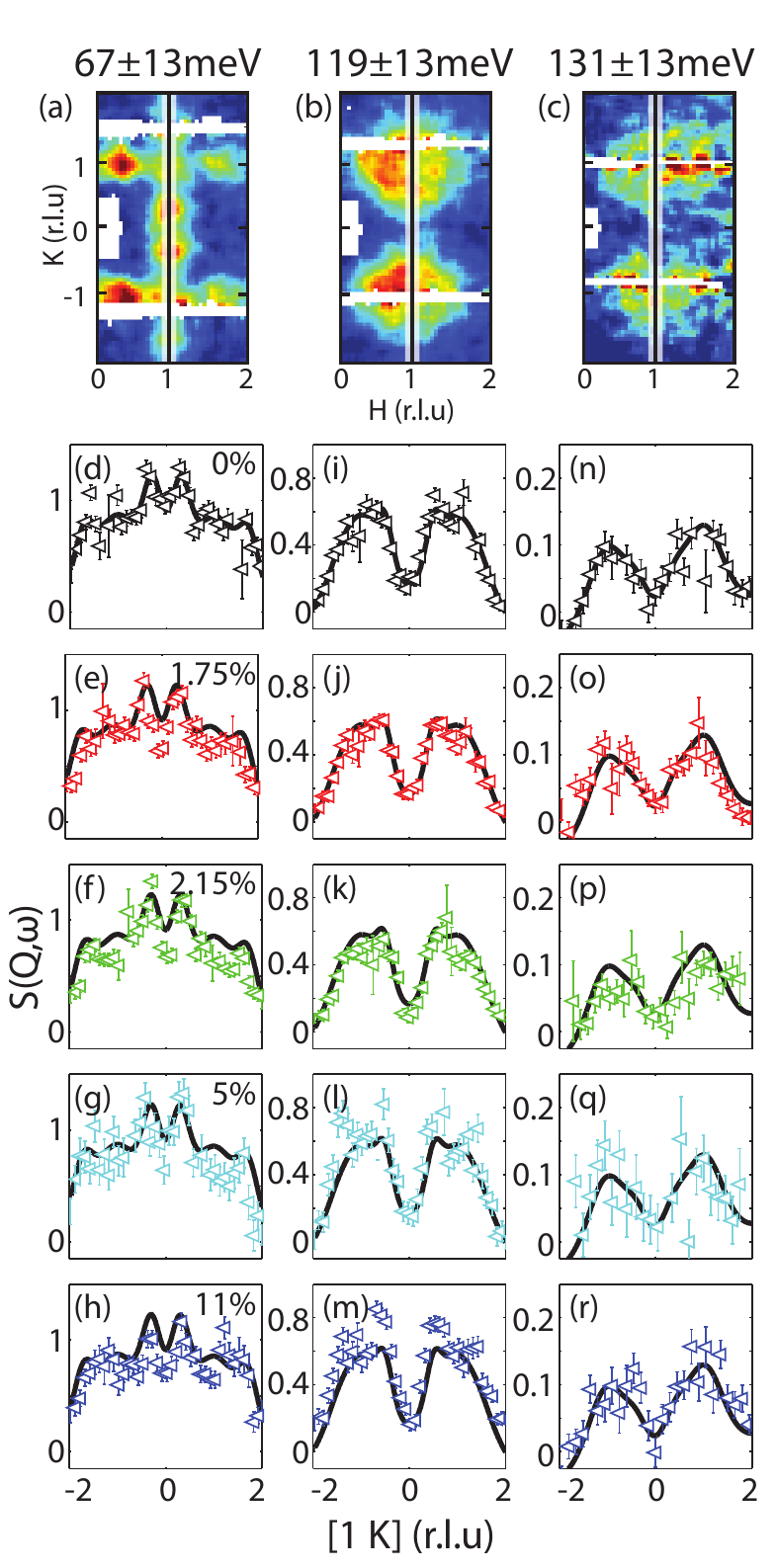}
	\caption{
		(Color online) 
		Illustrations of one dimensional transverse cuts from Fig. 4(a), 4(f), 4(k). Cuts measured with incident energy of 250meV at energy transfers of 67$\pm$13 meV (d)-(f), 119$\pm$13 meV (i)-(m), 171$\pm$13 meV (n)-(r) for doping values $x=0, 0.0175, 0.0215, 0.05$, and 0.11, respectively. All cuts integrated over $0.9<H<1.1$. Black line in each row is a Gaussian fitting to the parent compound for peak comparison.
	}
\end{figure}

We begin by examining the wave vector dependence of the two-dimensional (2D) background subtracted 
spin excitation intensities at different energy transfers 
as a function of increasing Co-doping $x$. 
Figure 2 summarizes the data acquired with an incident energy $E_i = 80$ meV at different excitation energies within the $[H,K]$ plane. The horizontal rows are excitation energies of $E=12\pm 4$, $28\pm4$, and $52\pm4$ meV, where the $\pm$ values indicate the range of energy integration. The columns show data from $x=0, 0.0175, 0.0215, 0.05$, and 0.11 in NaFe$_{1-x}$Co$_x$As. The white box in Fig. 2(a) illustrates the zone over which the magnetic scattering was integrated to estimate the local dynamic susceptibility $\chi^{\prime\prime}(E)$.  As expected, spin excitations are centered around AF ordering wave vectors ${\bf Q_{AF}}=(\pm 1,0)$ and $(0,\pm 1)$. With increasing Co-doping, $E=12\pm 4$ meV spin excitations at ${\bf Q_{AF}}=(1,0)$ become broader and weaker, and disappear completely for the $x=0.11$ nonsuperconducting sample [Figs. 2(a)-2(e)]. Upon increasing the excitation energies to $E=28\pm 4$ meV and $52\pm 4$ meV, the situation is similar except that spin excitations now appear for the $x=0.11$ sample [Figs. 2(f)-2(j), and 2(k)-2(o)].  Figure 3 summarizes one-dimensional cuts of the data along the transverse direction as illustrated in panels (a), (b), and (c). Black curves in Fig. 3 show single Gaussian fits to the excitations in the parent compound NaFeAs ($x=0$) and are over-plotted with cuts from $x=0.0175, 0.0215, 0.05$, and 0.11 for comparison of scale and peak width.  At the lowest probed energy transfer of $E=12\pm 4$ meV [Fig. 3(d)-3(h)], it is clear that there is a decrease of peak height and a small increase in peak width, with scattering persisting throughout the superconducting dome and fully gapped in the overdoped $x=0.11$ compound. This trend persists at $E=28\pm 4$ meV [Fig. 3(i)-3(m)] and becomes less apparent at $E=52\pm 4$ meV [Fig. 3(n)-3(r)]. 
At $E=52 \pm 4$ meV, the differences between different Co-dopings in terms of peak height, width, and splitting are very small. Large discrepencies in magnetic scattering intensity are observed only to small energy transfers with intensities becoming comparable around $E\approx 50$ meV. This observation is reflected in previous measurements in BaFe$_{2-x}$Ni$_x$As$_2$, in which strong doping dependence is observed below $E\approx 80$ meV \cite{msliu12,hqluo12,mwang13,hqluo13}.

Background subtracted constant energy transfer images of excitations for $E_i=250$ meV are shown in Fig. 4, with columns reflecting cobalt doping value in the same way as Fig. 2. Constant energy slices at $E=67\pm 7$ meV reveal very small changes in intensity and line shape with increasing cobalt doping [Fig. 4(a)-(e)]. The transverse dispersion from the AF ordering wave vector is clearly visible at all dopings, as is a feature stemming from the zone boundary wave vector positions $(\pm 1,\pm 1)$.  When energy is increased to $E=119\pm 7$ meV, spin excitations form a ring like structure centered around the wave vector $(\pm 1,\pm 1)$, and again are weakly Co-doping dependent [Fig. 4(f)-4(j)]. Upon increasing energy to $E=171\pm 7$ meV, spin excitations for all Co-doping levels become very weak but are well centered around  $(\pm 1,\pm 1)$ [Fig. 4(k)-4(o)].

Figures 5(d)-5(h) show transverse cuts along the $[0,K]$ direction at $E=67\pm 13$ meV [Fig. 5(a)], which reveal dispersive spin excitations away from the AF wave vector ${\bf Q_{AF}}=(1,0)$ and additional scattering at $(1,\pm 1)$. This additional scattering does not vary with doping, indicating it to be an intrinsic property of the NaFe$_{1-x}$Co$_x$As system. The presence of spin excitations near both wave vectors, ${\bf Q_{AF}}=(1,0)$ and $(1,1)$, is a unique feature of NaFe$_{1-x}$Co$_x$As not observed in BaFe$_{2-x}$Ni$_x$As$_2$ family of materials \cite{lharriger,hqluo13}. However, recent ToF INS measurements of the FeSe family of materials reveal spin excitations at both of these wave vectors that are interpreted as competition between stripe magnetic order and N$\rm \acute{e}$el order \cite{QSWang}.The presence of similar features in the NaFe$_{1-x}$Co$_x$As family of materials suggests that magnetic frustrations may also play an important role in determining the rather low N$\rm \acute{e}$el temperature and weakly ordered moment of the undoped NaFeAs \cite{SLi09a}. 
\begin{figure}[t]
	\includegraphics[scale=0.85]{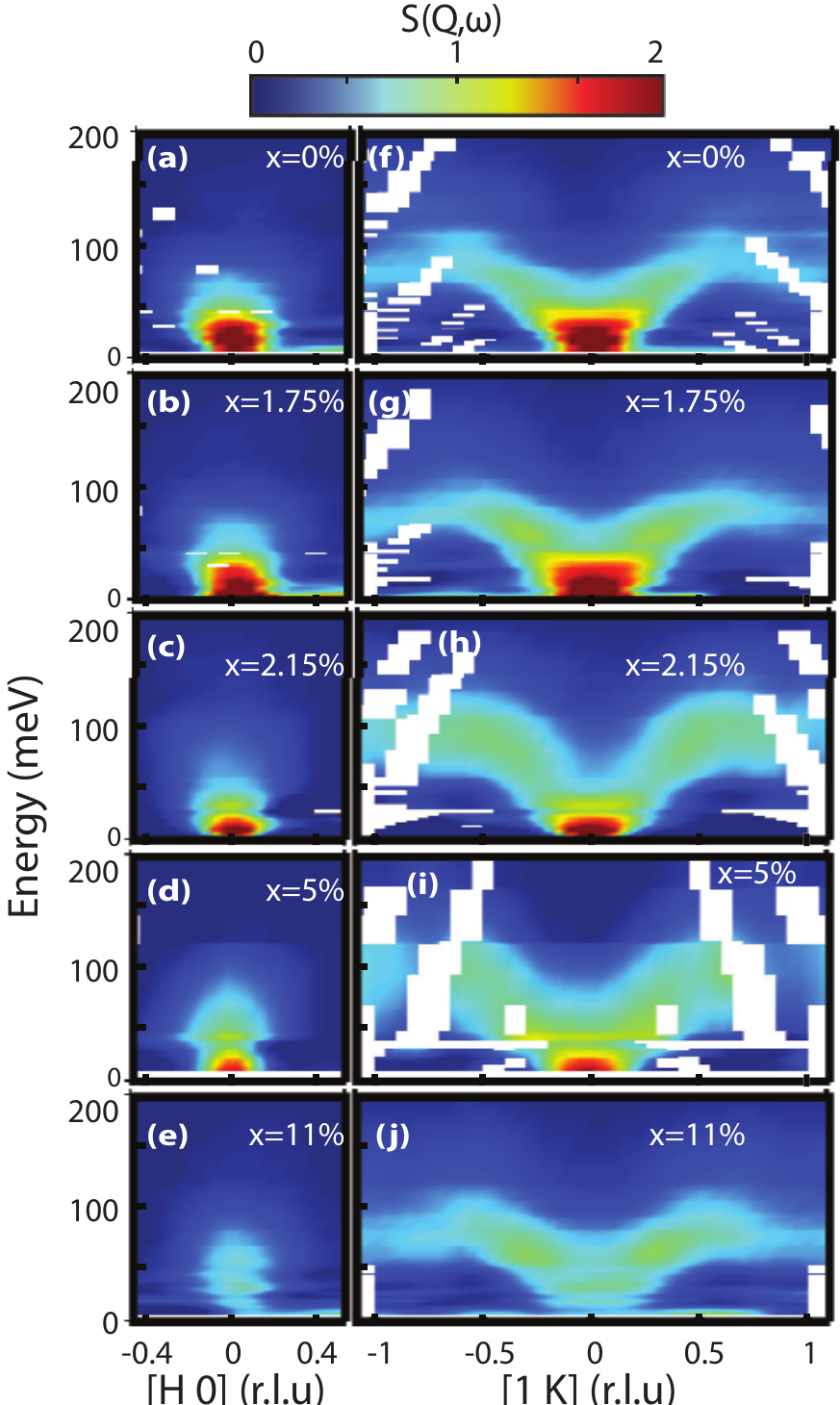}
	\caption{
		(Color online)
		(a)-(e) Two dimensional Q-E slices of NaFe$_{1-x}$Co$_x$As for $x =$  0, 0.0175, 0.0215,  0.05, and 0.11 , along H integrated on the window K = [-0.1 0.1] r.l.u. in absolute units. (f)-(j) Similar slices along K for $x =$ 0, 0.0175, 0.0215, 0.05, and  0.11  integrated across the window K = [0.9 1.1] r.l.u.. All figures use representative data from all incident energies measured and have been smoothed once using nearest neighbor points.
	}
\end{figure}

\begin{figure*}
	\centering
	\includegraphics[scale=0.7]{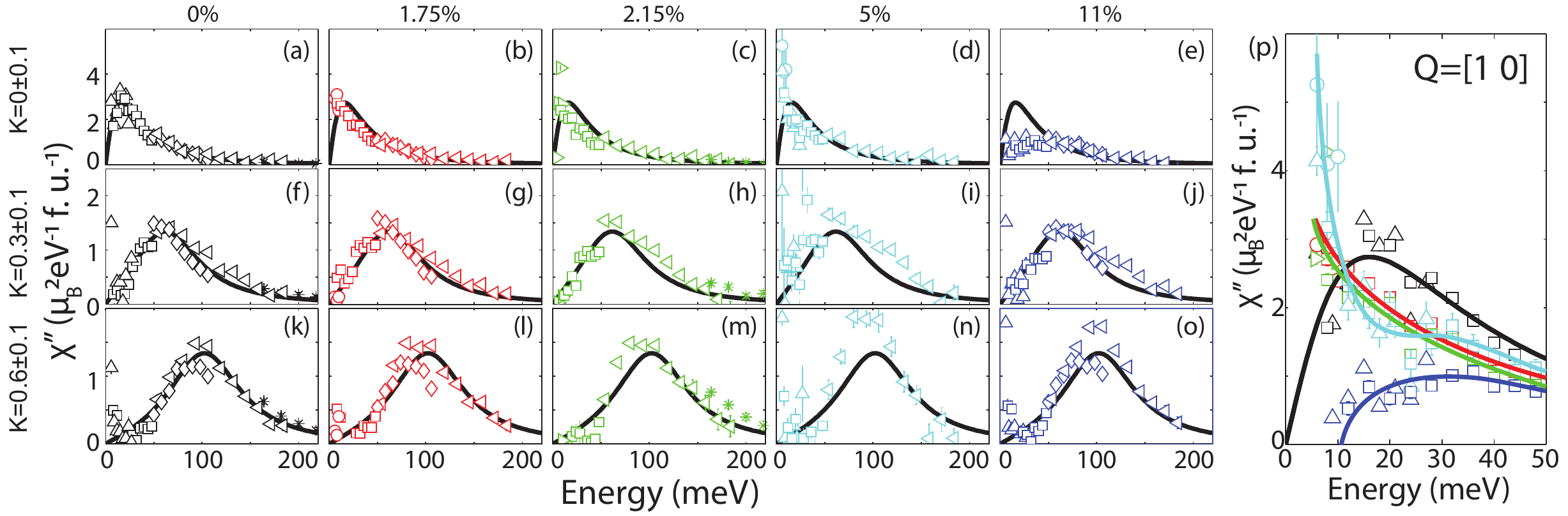}
	\caption{
		(Color online)
		One dimensional constant Q cuts obtained by integrating along H = [0.9 1.1] and (a-e) K = [-0.1 0.1], (f-j) K=[0.2 0.4], and (k-o) K=[0.5 0.7] for dopings of $x=0, 0.0175, 0.0215, 0.05$, and 0.11, respectively. Black lines are damped harmonic oscillator fits to parent compound data for comparison. Symbols correspond to incident energy (See Table II). (p) Overplot of low energy constant Q points at the AF wave vector with integration windows of ±0.1 in both directions. Solid lines are guides to the eye.
	}
\end{figure*}

At higher energy transfers, the signal becomes increasingly diffuse shown as broad peaks centered around $(1,\pm 1)$ at $E=119\pm 13$ meV [Figs. 5(i)-5(m), 5(b)] and  $E=171\pm 13$ meV [Figs. 5(n)-5(r), 5(c)]. These observations are broadly consistent with results in BaFe$_{2-x}$Ni$_x$As$_2$ \cite{hqluo13}. However, whereas spin excitations in the present compounds are already quite diffuse and centered at the zone boundary by 171 meV [Fig. 5(n)-5(r)] indicating the band top of a Heisenberg system, excitations at comparable energies in BaFe$_{2-x}$Ni$_x$As$_2$ remain well defined, indicating a smaller total magnetic excitation bandwidth in the NaFe$_{1-x}$Co$_x$As system. The high energy spin excitation intensities, lineshapes, and 
bandwidth are essentially Co-doping independent to the accuracy of our measurements.

To further study the effect of Co-doping on the overall magnetic excitation energy bandwidth, in Figure 6 we plot Co-doping dependent projections of the overall spin excitations along the $[1,K]$ and $[H,0]$ directions obtained using $E_{i}=250$ meV. Each figure is a compilation of background subtracted one-dimensional cuts, some of which are featured in Figs. 3 and 5.

We will first examine the transverse spin excitation dispersions along the $[1,K]$ direction for
different Co-doping concentrations.  Inspection of the data in Figs. 6(f)-6(j) reveals that 
the most obvious change with doping is in the low energy fluctuations. With increased Co-doping, there is a clear, systematic reduction of scattering intensity below 50 meV. Upon reaching the Co-overdoped state when superconductivity is suppressed, no magnetic scattering intensity is visible at small energy transfers due to the presence of a spin gap. 
For spin excitations above 50 meV, there is no distinct trend with increasing Co-doping, suggesting that 
the effective magnetic exchange coupling constants are weakly Co-doping dependent. 
Figures 6(a)-6(e) plot Co-doping dependence of the spin excitations projected along the $[H,0]$ direction.
Similar to data in Figs. 6(f)-6(j), low-energy spin excitations decrease with increasing Co-doping, and vanish for nonsuperconducting 
NaFe$_{0.89}$Co$_{0.11}$As. Additionally, while the dispersion is quite clear along the transverse $[1,K]$ direction, there is no dispersive feature longitudinally along the $[H,0]$ direction, illustrating the strongly anisotropic nature of the excitations.  These features are rather different from traditional spin waves from a local moment Heisenberg Hamiltonian.

To further compare the Co-doping evolution of the low-energy spin excitations,  
we plotted energy dependence of the dynamic susceptibility $\chi^{\prime\prime}(E)$ obtained by integrating the $E_i=250$ meV data over the range $H=1\pm0.1$ and $K=0\pm 0.1$ around the AF ordering wave vector ${\bf Q_{AF}}=(1,0)$ [Figs. 7(a)-7(e)].
Figures 7(f)-7(j) and 7(k)-7(o) show identical cuts ${\bf Q}$-integrated over   
$[H=1\pm0.1,K=0.3\pm 0.1]$ and $[H=1\pm0.1,K=0.6\pm 0.1]$, respectively. 
The solid lines in the figures are fits to undoped NaFeAs using the damped harmonic oscillator description of excitations $\chi^{\prime\prime}(E)=A \frac{\Gamma EE_0}{(E^2-E_0^2)^2+4(\Gamma E)^2}$, and overplotted on equivalent cuts of other Co-doping concentrations. While energy dependence of the dynamic susceptibility is weakly Co-doping dependent in Figs. 7(f)-7(j) and 7(k)-7(o), the low-energy $\chi^{\prime\prime}(E)$ clearly changes with increasing $x$ in Figs. 7(a)-7(e).  Figure 7(p) shows a magnification of the low energy excitations integrated around  
${\bf Q_{AF}}=(1,0)$, which clearly reveal a reduction of the low-energy dynamic susceptibility for NaFe$_{1-x}$Co$_x$As with $x=0.11$. 

In previous systematic INS studies of the hole/electron doping dependence of spin excitations in the BaFe$_2$As$_2$ family of materials \cite{msliu12,hqluo12,gstucker12,mgkim13,mwang13,hqluo13}, it was argued that the wave vector dependence of low-energy spin excitations arises from nesting between the hole and electron Fermi surfaces.  As Co- or Ni-doping to BaFe$_2$As$_2$ enlarges the electron Fermi surfaces and reduces tho hole Fermi surfaces \cite{richard}, spin excitations become transversely elongated following the doping-induced 
mismatch between the hole and electron Fermi surfaces \cite{JHZhang}. 
Figure 8 summarizes the full-width at half-maximum (FWHM) of spin excitations 
along the longitudinal [Fig. 8(a) inset and closed symbols in Fig. 8(a)-8(e)] and transverse [Fig. 8(a) inset and open symbols in Fig. 8(a)-8(e)] directions resulting from single Gaussian fits on data below $E=50$ meV. 
A direct comparison across dopings can be seen in Fig. 8(f). 
While solid lines are for the longitudinal FWHM and overlay well, 
the transverse elongation, shown in dashed lines, increases with increasing Co-doping below $E\approx 40$ meV. 
For comparison, the transverse FWHM for optimally doped BaFe$_{1.9}$Ni$_{0.1}$As$_2$ 
is plotted in Fig. 8(a) as the dashed line \cite{msliu12}. At low energy transfers, the FWHM is the same but grows more slowly with energy due to a higher spin excitation velocity.

In order to quantitatively compare the Co-doping dependent dispersion curves of NaFe$_{1-x}$Co$_x$As, we plot in Figure 9 the evolution of spin excitation dispersions along the $[1,K]$ direction as a function of increasing Co-doping $x$.   
 Open symbols are fits from two Gaussion fits to transverse constant energy cuts through the AF wave vector. 
Filled symbols are peak centers from constant wave vector cuts like those shown in Fig. 7. 
Consistent with data in Fig. 6, we find that Co-doping into NaFeAs broadens the low-energy spin excitations along the transverse direction, but has little impact to the overall dispersion or spin excitation bandwidth of the system.  This is most clearly illustrated in 
Fig. 9(f), where dispersions for different Co-dopings are overplotted.  

\begin{figure}[t!]
	\includegraphics[scale=.78]{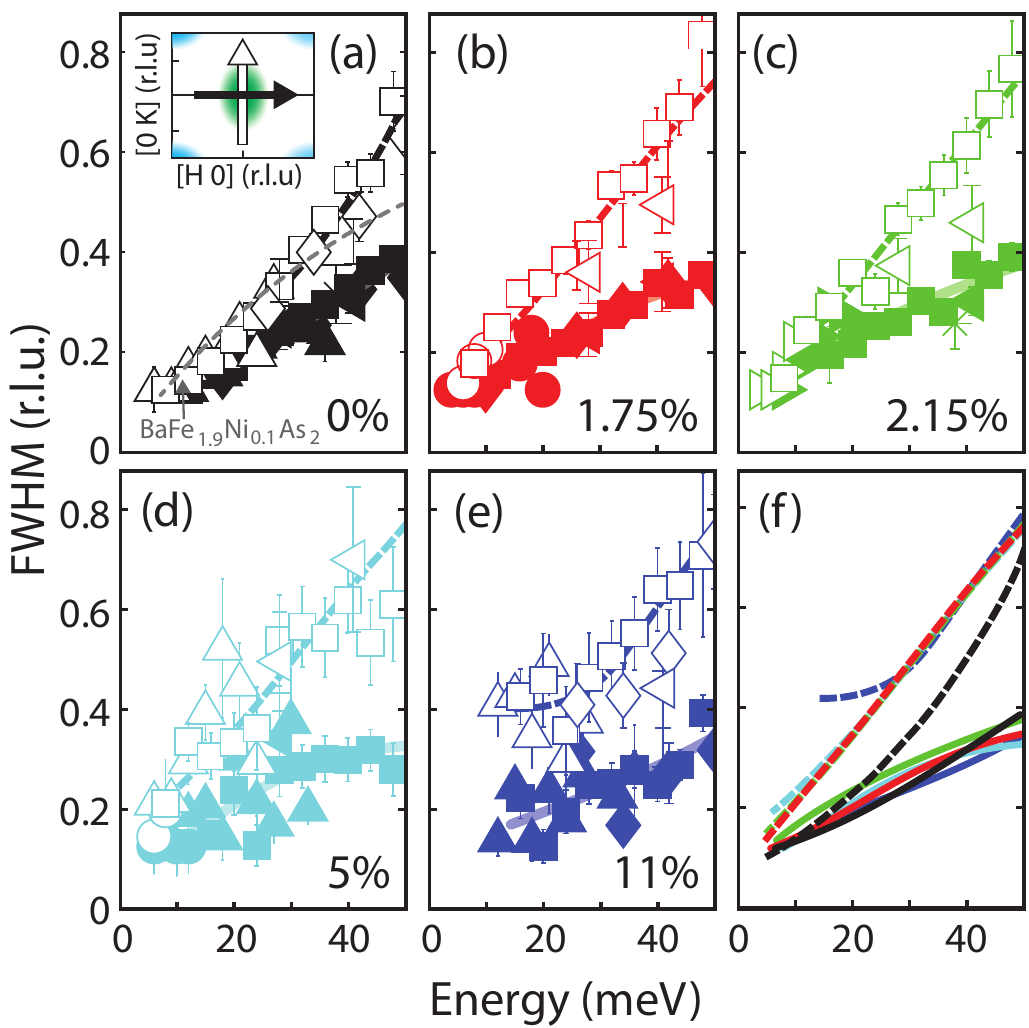}
	\caption{ (Color online) 
		(a) [Inset] Schematic showing direction of cuts at the AF wave vector [1 0]. Solid arrow (and points) indicates a longitudinal cut and open arrow (points) indicates a transverse cut. (a-e) FWHM of single Gaussian peak fits to constant energy cuts. Dashed and solid lines are guides to the eye for longitudinal and transverse cuts respectively. (f) Overplot of guides from (b-f).
	}
\end{figure}

\begin{figure}[t]
	\includegraphics[scale=0.78]{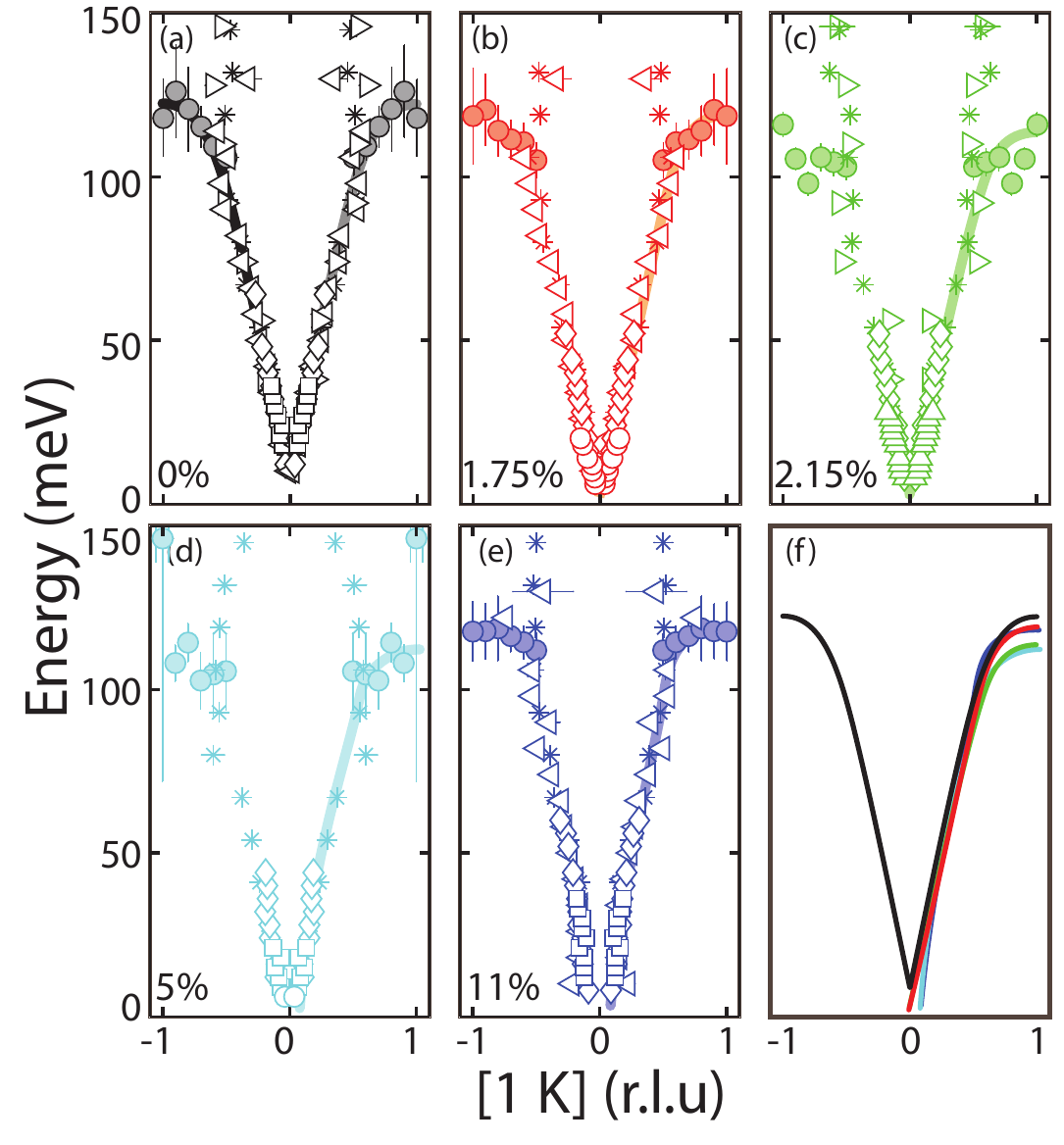}
	\caption{ (Color online) 
		(a-e) Dispersion along the [1 K] direction from two-gaussian fits to constant energy cuts (open symbols) and damped harmonic oscillator fits to constant energy cuts (filled symbols) for $x = 0, 0.0175, 0.0215, 0.05$, and 0.11, respectively. Solid lines are guides to the eye. (f) Overplot of guides from (a-e). 
	}
\end{figure}

\begin{figure}[t]
\includegraphics[scale=1]{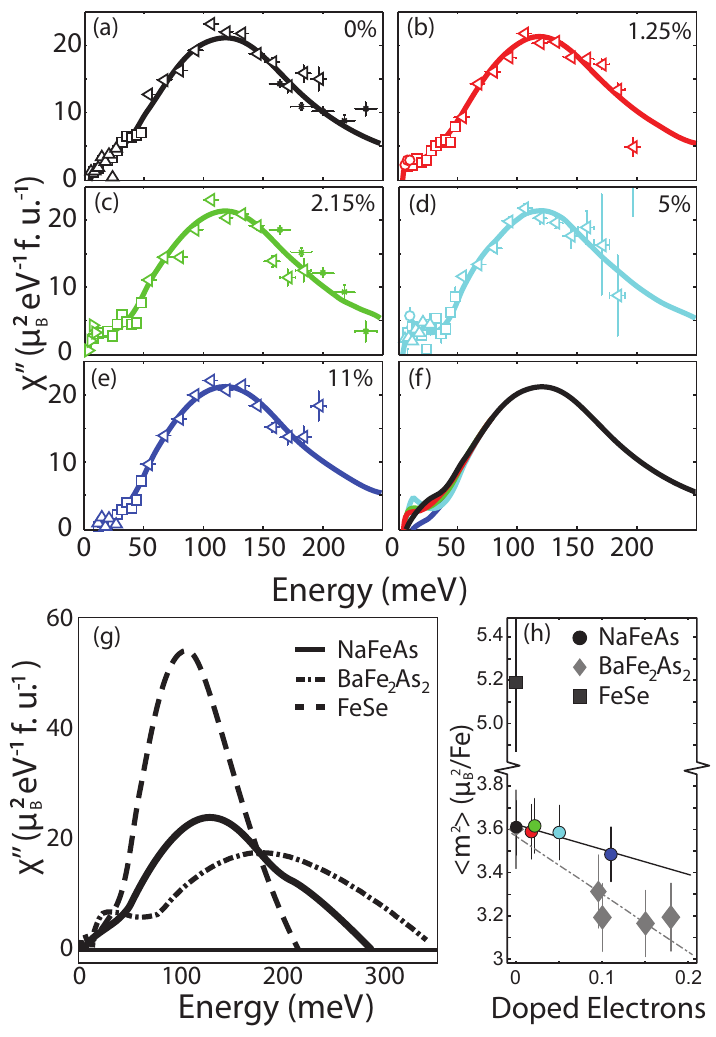}
\caption{ (Color online) 
Energy dependence of the local dynamic susceptibility $\chi^{\prime\prime}(E)$ in NaFe$_{1-x}$Co$_x$As in absolute unites for (a-e) $x = 0, 0.0175, 0.0215, 0.05$, and 0.11, respectively. Solid lines area combination of a smoothed moving average below E = 70meV and a damped harmonic oscillator fit to all data sets above 70 meV. (f) Overplot of guides from (a-e). (g) Energy dependence of the local susceptibility in the parent compounds of NaFeAs (solid line), BaFe$_2$As$_2$ (dashed-dotted line), and FeSe (dashed line).  (h) Dependence of total fluctuating moment on doped electrons in NaFe$_{1-x}$Co$_x$As (circles), BaFe$_{2-x}$Ni$_x$As$_2$ (diamonds), and FeSe (square). Solid line is a linear fit to the electron doping dependency of the moment in NaFe$_{1-x}$Co$_x$As. Dashed line is an identical fitting for BaFe$_{2-x}$Ni$_x$As$_2$
}
\end{figure}

Figures 10(a)-10(e) illustrate the energy and doping dependence of the local dynamic susceptibility $\chi^{\prime\prime}(E)$ for different Co-doping concentrations.  The wave vector integration range of $\int\chi^{\prime\prime}(E,{\bf Q})dQ/\int dQ$ is shown in the dashed box of Fig. 2(a) \cite{dai15}.  Solid lines in the figures are a combination of a guide to the eye derived from a moving average of data for energy transfers below 70 meV and a damped harmonic oscillator fit to data from all dopings above 70 meV energy transfer. The horizontal error bars indicate energy integration range, and the vertical error bars are statistical errors from the integration.
Figure 10(f) plots the solid lines from 10(a)-10(e), which reveal that  $\chi^{\prime\prime}(E)$ for energies above 70 meV are virtually identical at all probed Co-doping levels.  To compare these results with those obtained for the BaFe$_2$As$_2$ and the FeSe families of iron-based superconductors, we show in Fig. 10(g) energy dependence of $\chi^{\prime\prime}(E)$ for FeSe (dashed line) \cite{QSWang}, BaFe$_2$As$_2$ (dashed dotted line) \cite{Harriger12}, and NaFeAs (solid line) \cite{CLZhang14}.  It is clear that spin excitation energy bandwidth systematically decreases on moving from BaFe$_2$As$_2$ to NaFeAs and then to FeSe.  This is consistent with the notion that electron correlations increase from BaFe$_2$As$_2$ to NaFeAs, then to FeSe due to the increased iron pnictogen height from the iron plane \cite{ZPYin11,ZPYin14}.

To determine the total fluctuating magnetic moments of NaFe$_{1-x}$Co$_x$As, defined as $\left\langle m^2\right\rangle=(3/\pi)\int\chi^{\prime\prime}(E)dE/(1-\exp(-E/kT))$ \cite{clester10}, and compare the outcome with those of BaFe$_{2-x}$Ni$_x$As$_2$ \cite{hqluo13} and FeSe \cite{QSWang}, we show in Fig. 10(h) the electron-doping dependence of $\left\langle m^2\right\rangle$ for the first two families of materials, where the electron doping level per iron is assumed to be Co-doping (or 1/2 Ni-doping) level per Fe site. By overlaying the electron doping dependence of the total fluctuating magnetic moments from all compounds, we see a systematic decrease in $\left\langle m^2\right\rangle$ with increasing electron doping. In spite of the largely different energy scales of the overall spin excitation bandwidth for BaFe$_{2-x}$Ni$_x$As$_2$, NaFe$_{1-x}$Co$_x$As, and FeSe, their total fluctuating magnetic moments are rather similar, and decrease systematically with the number of electrons added rather than irons replaced.  These results reinforce the view that magnetism is important for superconductivity of iron-based superconductors regardless of how electrons are doped into these materials \cite{scalapino,dai}.  

\begin{figure}[t]
\includegraphics[scale=1]{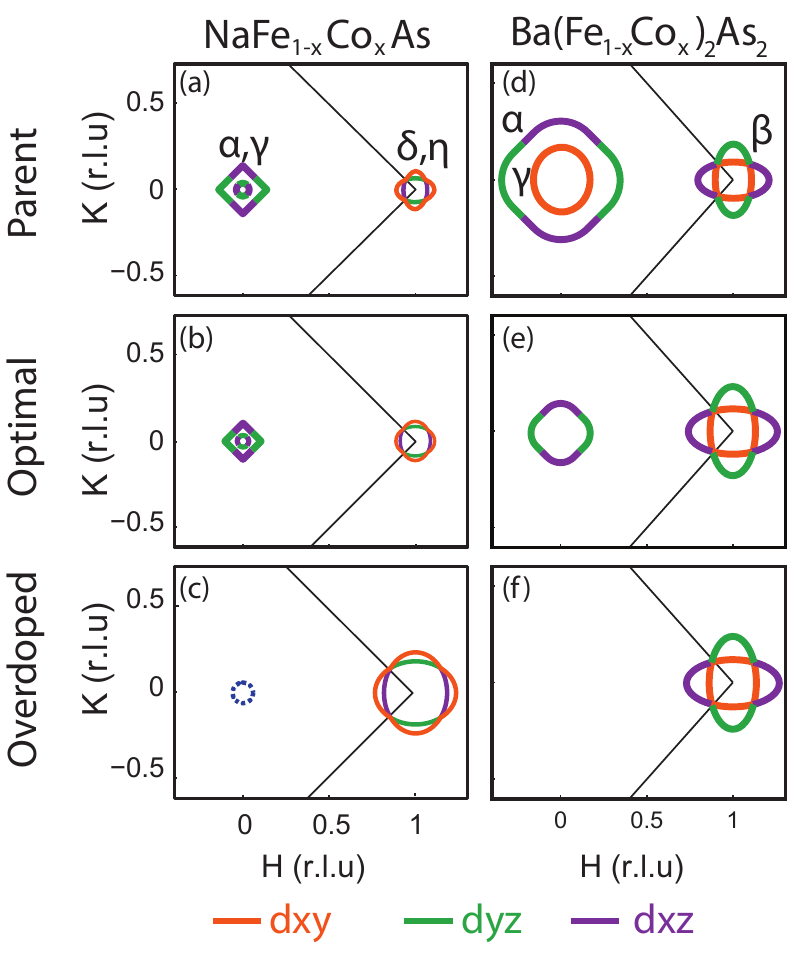}
\caption{ (Color online) 
Schematic Fermi surface for NaFe$_{1-x}$Co$_x$As and Ba(Fe$_{1-x}$Co$_x$)$_2$As$_2$ in the (a,d) parent\cite{CHea,dai15}, (b,e) optimally doped\cite{ZRYe,dai15}, and (c,f) overdoped\cite{STCui12,dai15} regimes respectively. ARPES results show the $\alpha$ and $\gamma$ bands are hole-like whereas $\beta$ and $\eta$ are electron bands. Dashed lines indicate a band is very near but just below the Fermi Surface.
}
\end{figure}

\section{Discussion and Conclusions}

Through a comprehensive survey of spin excitations in NaFe$_{1-x}$Co$_x$As, we establish electron doping evolution of the spin excitation spectra for this family of iron pnictide superconductors. 
Figure 11 compares the electron doping evolution of the Fermi surfaces of NaFe$_{1-x}$Co$_x$As \cite{ZHLiu11,STCui12,QQGe13} and Ba(Fe$_{1-x}$Co$_x$)$_2$As$_2$\cite{richard} obtained from angle resolve photoemission spectroscopy experiments, where the red, green, and purple colors mark $d_{xy}$, $d_{yz}$, and $d_{xz}$ orbital characters of the Fe $3d$ electrons.  In spite of the clear differences in the Fermi surfaces of undoped NaFeAs and BaFe$_2$As$_2$, the effects of electron-doping by partially substituting Co for Fe are similar. Namely, the electron Fermi surfaces are enlarged and the hole Fermi surfaces are drastically reduced. Superconductivity vanishes when hole pockets near $\Gamma$ sink below Fermi surface due to electron over-doping, destroying the hole-electron Fermi pockets nesting condition. This reinforces the view that while the nesting condition is not sufficient for the presence of superconductivity, the correlation of its destruction with the disappearance of superconductivity suggests it is a necessary condition for some iron pnictides \cite{mazin2011n,Hirschfeld,Chubukov}.  Our systematic measurements of the overall spin excitation spectra in NaFe$_{1-x}$Co$_x$As family of materials are consistent with the picture where the low-energy spin excitations are coupled with the Fermi surface nesting condition while high-energy spin excitations are much less Co-doping dependent. The evolution of spin excitations in the BaFe$_{2-x}$Ni$_x$As$_2$ family of materials paints a very similar picture \cite{hqluo13}.  In addition, we find that in spite of the large differences in spin excitation bandwidth amongst the NaFe$_{1-x}$Co$_x$As, BaFe$_{2-x}$Ni$_x$As$_2$, and FeSe families of materials, their total spin fluctuating moments are comparable to within $\sim$50\% and decrease with increasing electron doping. This is surprising given that these families of iron-based superconductors have rather different crystal structures and ground states: BaFe$_2$As$_2$ has nearly coupled structural and magnetic phase transitions with a static ordered moment of $M=\sim$0.8 $\mu_B/$Fe \cite{qhunag}, NaFeAs has separated structural and magnetic phase transitions with $M=\sim0.1\ \mu_B/$Fe \cite{SLi09a}, and FeSe is a superconductor without static AF order \cite{FCHsu}. Their similar total spin fluctuating moments suggest that the microscopic origin for magnetism, and possibly also superconductivity, is the same for these materials. 

\begin{figure}[t]
	\includegraphics[scale=1]{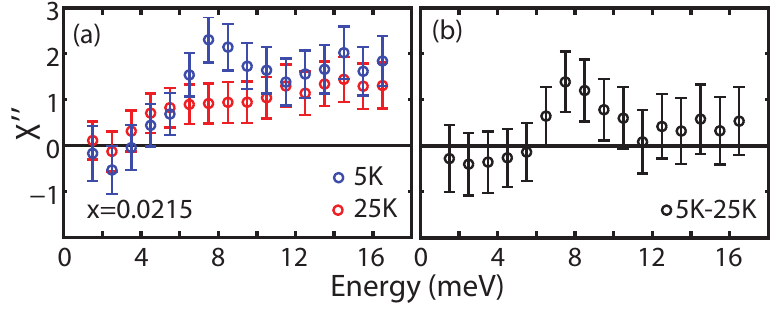}
	\caption{ (Color online) 
		(a) Background subtracted local susceptibility of NaFe$_{1-x}$Co$_x$As ($x=0.0215$) and (b) the difference in susceptibility between the superconducting ($T=5$ K) and normal ($T=25$ K) state. High temperature data are measured from the same samples and environment as other data for $x=0.0215$ in the present report.
	}
\end{figure}

It has been widely observed that there is a gain in low energy spin fluctuations, dubbed the neutron spin resonance, upon entering the superconducting state, suggesting a close connection between superconductivity and magnetic fluctuations. Quantitatively, it is reasonable to assume a connection if the energy gain from these fluctuations is larger than the superconducting condensation energy. This was shown to be the case in Ba$_{0.67}$K$_{0.33}$Fe$_2$As$_2$\cite{mwang13} where an exchange energy $\Delta E_{ex} = -0.66$ meV/Fe was much larger than the condensation energy $U_c=$-0.09 meV/Fe. Following the same procedure, we consider if the same is true in near optimally doped NaFe$_{1-x}$Co$_x$As. The local susceptibility above and below $T_c$, shown in Fig. 12(a), gives rise to an exchange energy of $\Delta E_{ex} = -0.21$ meV/Fe, while the condensation energy, calculated from the specific heat of a similarly doped compound \cite{GTTan13}, was found to be $U_c$=-0.008 meV/Fe. While these compounds have similar exchange energies, the superconductivity in NaFe$_{1-x}$Co$_x$As is substantially more fragile, and is several orders of magnitude smaller than the exchange energy associated with resonance. Therefore, our results are consistent with the notion that spin excitations are responsible for superconductivity in NaFe$_{1-x}$Co$_x$As.

In conclusion, using ToF INS spectroscopy, we have mapped out the overall spin excitations spectra for 
 NaFe$_{1-x}$Co$_x$As with $x=0,0.0175,0.0215,0.05,0.11$.  Our central conclusion is that the electron-doping evolution of the spin excitations spectra in this family of iron pnictides is similar to those of electron-doped BaFe$_2$As$_2$ family of materials, in spite of their large differences in structure and total magnetic excitation energy bandwidth. Given the similarities present across different families of iron pnictides, our data suggest they share a microscopic origin for magnetism and superconductivity and highlight the coupling between their spin excitations superconductivity \cite{scalapino}. 

\section{Acknowledgments}

A special thanks to Zachary Simms, Tucker Netherton, and Caleb Redding for great contributions in synthesiss. The single crystal synthesis and neutron scattering work at Rice University 
is supported by the U.S. DOE, Office of Basic Energy Sciences, under
Contract No. DE-SC0012311. Part of the
materials synthesis work at Rice University is supported by the Robert
A. Welch Foundation Grant No. C-1839. The research
at ORNL's SNS and HFIR was sponsored by
the Scientific User Facilities Division, Office of Basic Energy Sciences, U.S.
DOE.

\section{Appendix}

\subsection{Absolute Neutron Scattering Intensity Normalization}

Typically, ToF INS data is normalized using a vanadium standard for samples with known mass in the neutron beam. The NaFe$_{1-x}$Co$_2$As system, however, is more difficult to normalize by the sample mass as there is often residual powder flux trapped in the single crystal sample   
during its formation, especially at high Co-doping concentrations. 
This fact makes normalization purely by vanadium unreliable as the single crystal mass contributing to coherent scattering may be lower than the weighed mass. Because we wanted to compare spin excitation intensities directly across several dopings and with other iron-based superconductors, we sought self-consistent normalization which could be checked against an external standard. This requires a comparison of the structural properties rather than magnetic ones, and due to the varied incident energies measured for each composition (Table I) as well differing detector geometries, there were not many viable options. We identified one optical phonon as well as one acoustic phonon usable for self-consistent normalization and one acoustic phonon used for absolute normalization. All data shown in the appendix from ARCS ($x=0, 0.0175$, and 0.11) had been normalized to the same vanadium sample and 
data from SEQUOIA ($x=0.0215$) had also been normalized to a standard vanadium.
\begin{figure}[!t]
	\includegraphics[scale=0.45]{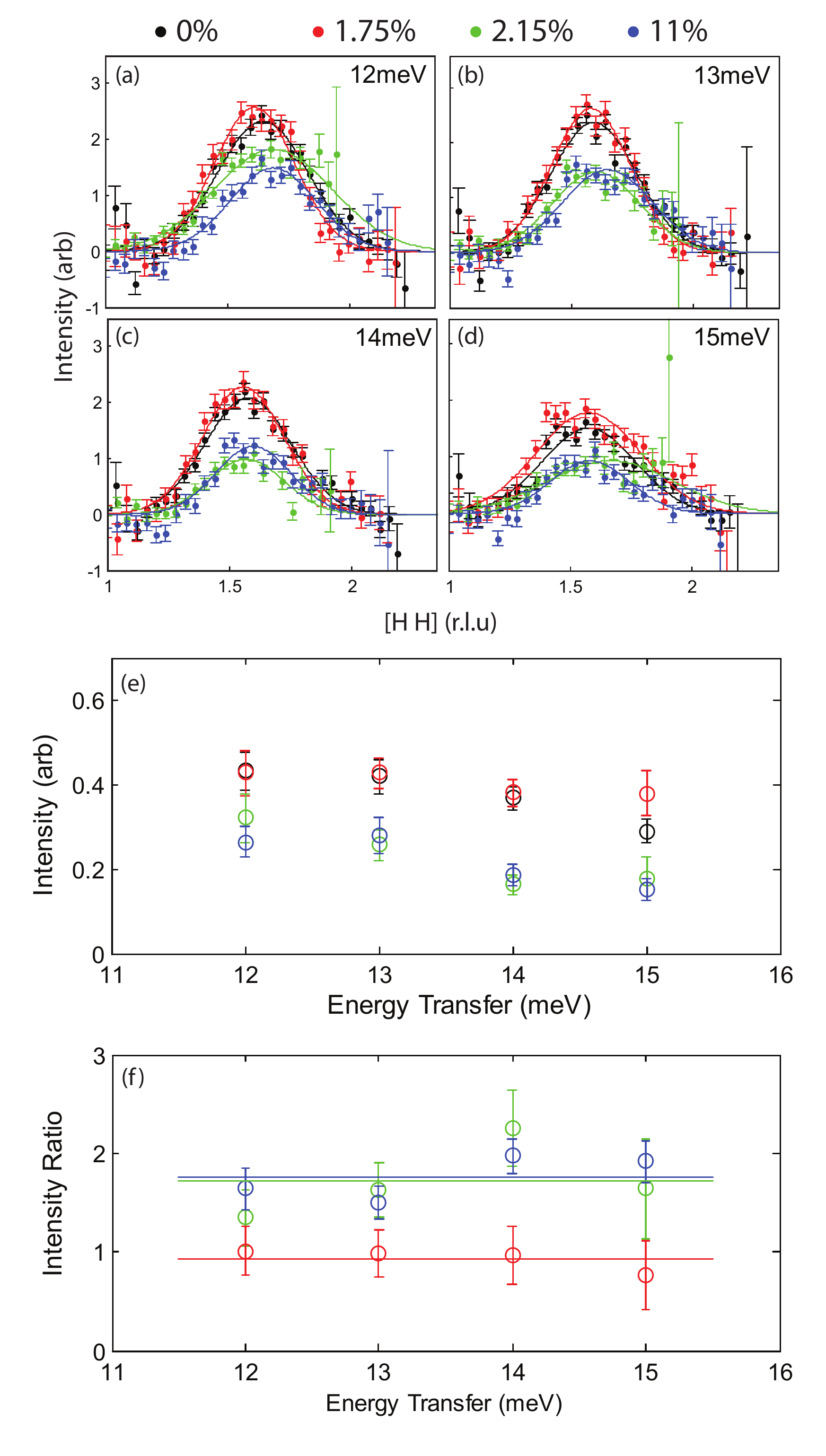}
	\caption{(color online) 
		Self-consistent normalization using the top of the $(2,2,1)$ acoustic phonon. (a-d) Background subtracted cuts along the $[H,H]$ direction at $E=12\pm 0.5, 13\pm0.5, 14\pm0.5, 15\pm0.5$ meV respectively. Dopings are noted by color and labeled at the top of the figure. Single peak Gaussian fits are plotted in solid lines. (e) Integrated intensity for cuts in (a-d) are plotted as a function of energy. Integrated intensity is calculated from fits with error bars deriving from fit parameters. (f) Intensity ratio ($I_{x=0}/I_x$) are plotted as a function of energy. Horizontal lines depict the weighted average of corresponding data.}
\end{figure}
\begin{figure}[!t]
	\includegraphics[scale=1]{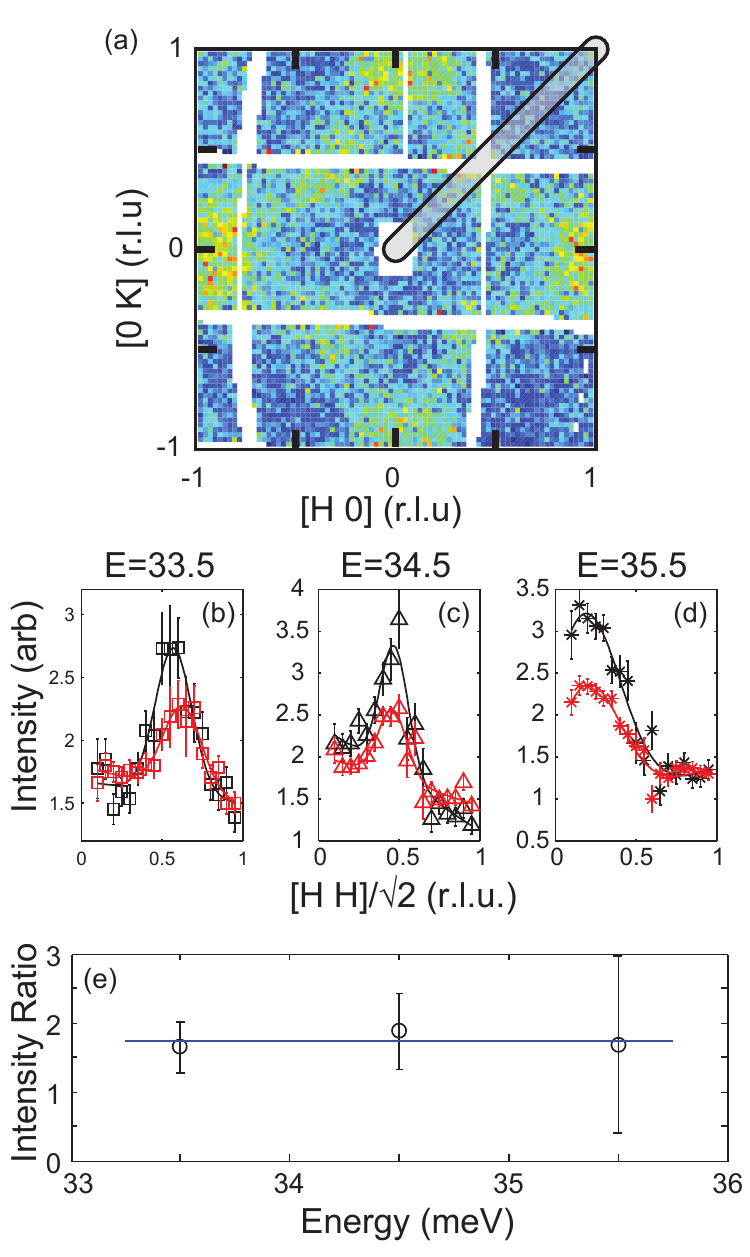}
	\caption{(color online) 
		(a) Optical phonon near the zone center seen with $E_i=50$ meV and $E=34.5\pm 0.5$ meV. (b-e) Cuts through the phonon for NaFe$_{1-x}$Co$_x$As with $x=0$ (black) and 0.11 (red). Cuts fit with single Gaussians (solid lines) constrained to share  a center. (e) Ratios of integrated intensity for fits in (b-d). Blue line is a weighted average.}
\end{figure}
\begin{table}[!b]
	\centering
	\caption{Normalization Factors}
	\begin{tabular}{c|c|c|c|c|c}
		Doping  & 0\% &    1.75\%     &    2.15\%     & 5\% &     11\%      \\ \hline
		Acoustic &  1  & 0.95$\pm$0.05 & 1.68$\pm$0.19 &  -  & 1.75$\pm$0.11 \\
		Optical  &  1  &       -       &       -       &  -  & 1.73$\pm$0.11 \\
		Absolute &  -  &       -       & 1.69$\pm$0.19 &  -  &       -       \\
		Given   &  -  &       -       &       -       & 1.9 &       -
	\end{tabular}
\end{table}

For $E_i=250$ meV, we found it possible to see the acoustic phonon near $(2, 2, 1)$ at the edge of the detector at ARCS as well as SEQUOIA. Unfortunately, the smaller detector area at MAPS prevented us from comparing the $x=0.05$ Co-doped compound. Cuts were made at 
$E=12\pm 0.5, 13\pm 0.5, 14\pm 0.5, 15\pm 0.5$ meV [Figs. 13(a)-13(d)] where intensity was strongest due to proper $L$ matching. Given only modest changes to lattice parameters and the similarity in mass between cobalt and iron, changes to terms such as the dynamic structure factor are small and ignored for this discussion. Comparing integrated intensities for the phonon at different energies [Fig. 13(f)], we arrived at a set of scaling factors for self-consistent normalization [Table S1].

The second check of normalization comes from data with $E_i=50$ meV for $x=0$ and 0.11 doped compounds. An optical phonon is clearly seen in the energy slices from $E=33$-36 meV moving toward $(0, 0)$ along the $[H, H]$ direction [Fig. 14(a)]. Making identical longitudinal cuts seen in Figure 14(b)-14(d) shows the phonon can be identified and fit with a single Gaussian curve. Using the fit parameters to calculate the intensity ratio for the overdoped compound results in an identical scaling parameter as the normalization by acoustic phonons. This gave us confidence in the scaling results.

\begin{figure}[!t]
	\includegraphics[scale=1]{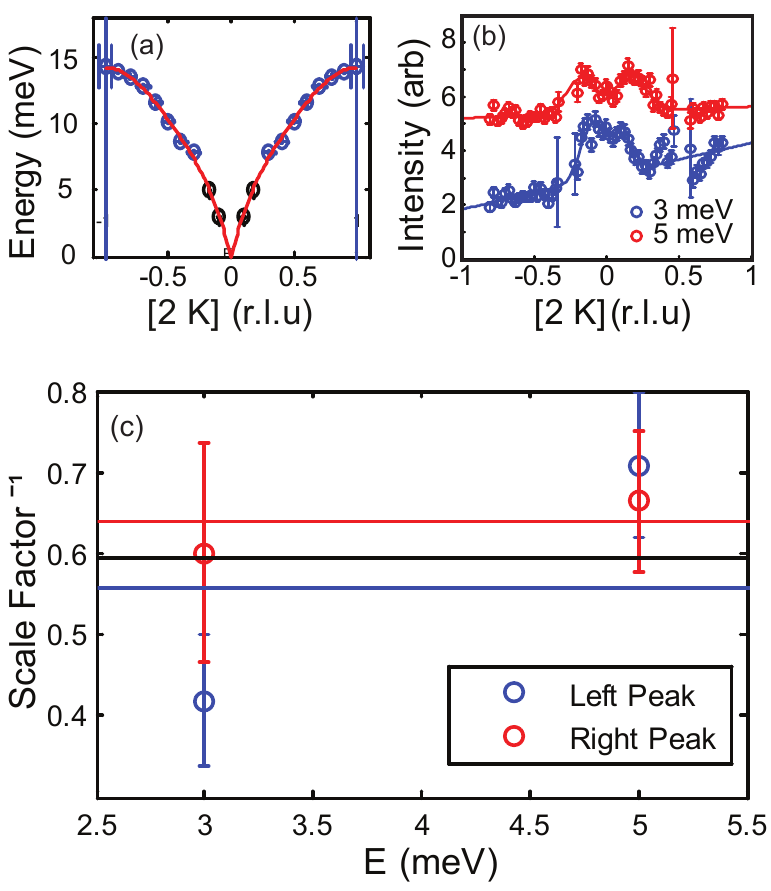}
	\caption{(color online) 
	(a) Acoustic phonon from $(2,0,1)$ determined by constant $E$ and Q scans. Red line is a polynomial fit used to determine the phonon speed. (b) Transverse cuts at 3$\pm$1 (lower) and 5$\pm$1 meV (higher) fit with two Gaussians on a linear background (solid lines). (c) Following ref\cite{PhonNorm}, the inverse scale factor resulting from self normalization with phonon shown in (a-b).}
\end{figure}

For an absolute certainty that these scaling results were reliable, we used a clearly visible phonon at $(2,0,1)$ in the $x=0.0215$ Co-doped compound with $E_i=35$ meV. The entire dispersion, mapped from constant ${\bf Q}$ and $E$ cuts in Figure 15(a), has different $L$ values at the peak positions. Figure 15(b) shows constant-energy cuts along the $[2,K]$ direction for
$E=3$ and 5 meV, which show clear counter propagating phonons. Figure 15(c) plots the scale factor 
obtained using different energy transfers.
Energy transfers near $E=4$ meV gave the proper value $L=1$ and were used in the self normalization. Using the process outlined by Xu {\it et al.} \cite{PhonNorm}, we found a self-normalization factor nearly identical to the one derived from the self-consistent normalization to the parent compound.

\subsection{Background Subtraction}
Once these scaling factors were taken into account, the high energy part of the local susceptibility overlapped as seen in Fig. 10(f). This brought all samples in line except for the $x=0.05$ Co-doped compound which, due to differing detector geometry, was not able to be included in any self-consistent normalizations. Given the universality of high energy excitations across the phase diagram, a scale factor was chosen for the $x=0.05$ cobalt doped sample to bring it in line. Final values for normalization scale factors can be found in Table I.
\begin{table}[!t]
	\centering
	\caption{Measured Incident Energy Summary for NaFe$_{1-x}$Co$_x$As}
	\begin{tabular}{|c|c|c|c|c|c|c|}
		\hline
		Ei (meV) & 0  & 0.0175 & 0.0215  & 0.05  & 0.11 & Symbol \\ \hline
		25    &  -   &   X    &    -    &  X   &  -   & \small{$\bigcirc$} \\
		35    &  -   &   -    &    X    &  -   &  -   & \Large{$\triangleright$} \\
		50    &  X   &   -    &    -    &  -   &  X   & $\bigtriangleup$ \\
		80    &  X   &   X    &    X    &  X   &  X   &  $\Box$\\
		150    &  X   &   X    &    -    &  -   &  X   &  $\lozenge$ \\
		250    &  X   &   X    &    X    &  X   &  X   & \large{$\ast$} \\
		350    &  X   &   -    &    X    &  -   &  -   & $\bullet$ \\ \hline
		Instr   & ARCS &  ARCS  & SEQUOIA & MAPS & ARCS & - \\ \hline
	\end{tabular}
\end{table}
When desiring to directly compare intensities of ToF INS data, it is of paramount importance that care is taken when subtracting the background. Background due to the sample environment can, in principle, be determined by measuring an empty environment without sample. We chose not to do this but instead focused our efforts on increasing counting statistics. This is reasonable considering a majority of the low energy background is not from aluminum, but rather phonons from the sample itself. Thus, the first challenge comes from carefully fitting and subtracting background due to low-energy phonons. Upon reaching higher energies, the spin excitations broaden as they disperse toward the zone boundary wave vectors. Near the band top, the excitations become broad and diffusive occupying a large fraction of the zone boundary, making it tricky to discern true intensity from background. Distinct energy regions we identified each require their own background fitting solution.
\begin{figure*}[!t]
	\centering
	\includegraphics[scale=1]{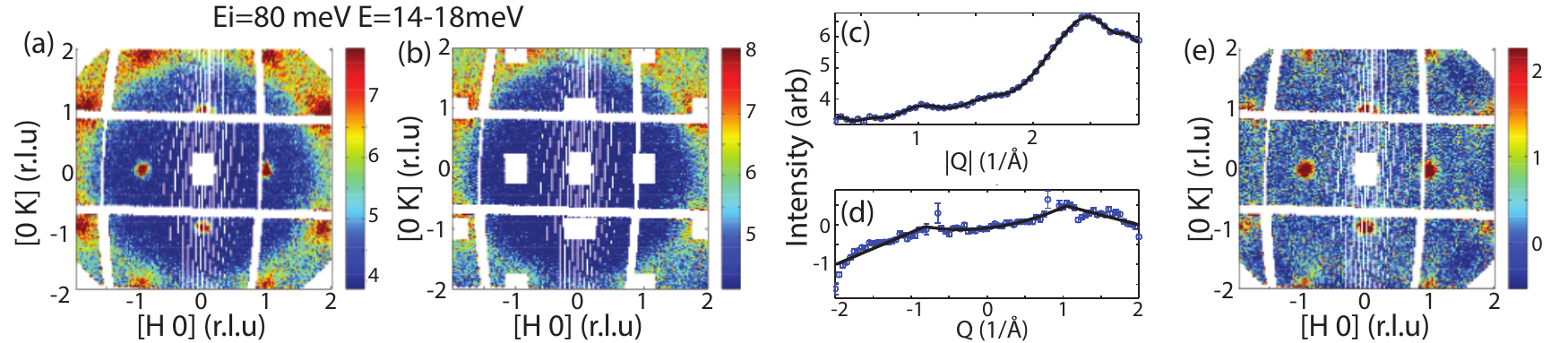}
	\caption{(color online) 
		Background subtraction process at low energies. Raw data (a) is masked according to the method outlined in the text (b). Ring-integrated radial data (c) and vertical linear background (d) is subtracted to give a background-free slice (e).}
\end{figure*}
At low energies, the background is predominantly due to phonons and detector quirks. It should be noted that all ToF measurements were performed in the same orientation, tying the $L$ component of ${\bf Q}$ to the incident and transferred energies. This is reasonable since triple axis measurements have revealed only a weak $L$-dependence in spin excitations of NaFe$_{1-x}$Co$_x$As \cite{CLZhang13b} removing the need to consider $L$ values. Given that the phonons are well defined in ${\bf Q}$ and $E$, measurements made at different incident energies will result in different phonon backgrounds. Fortunately, the phonon background in the vicinity of the AF wave vector is always quite broad. This is because the AF wave vector is at the edge of the structural zone boundary. so the dispersion is already quite flat when it reaches the AF wave vector. This broad nature, alongside the high symmetry of the twinned crystal, leads to nearly isotropic features.

An example of our background fitting and subtraction process is highlighted in Figure 16. Figure 16(a) shows the raw data with $E_i=80$ meV at energy transfer of $E=16\pm 2$ meV. 
we first mask signal at AF wave vectors shown in white boxes in Fig. 16(b).
we then fit a radial background by integrating rings of constant ${\bf Q}$ and $E$ after masking the data [Fig. 16(c)]. Intensity was fit using a high-order polynomial with order decreasing with energy. Masking was done by fitting transverse and longitudinal cuts and omitting data within three half-width at half-max of peak centers [Fig. 16(b)]. Additionally, we found a large background component along the vertical direction of detector tubes. In fact, each detector bank had a distinctly different profile. This may be due in part to the large asymmetry of our sample mount along the vertical direction. Background parallel to the detector tubes was also fit after subtracting the radially symmetric component [Fig. 16(d)]. A masked low-${\bf Q}$ region was used for this fitting. This method was used for energy transfers below $E=50$ meV. The background subtracted data is shown in Fig. 16(e), where we find clear magnetic excitations at the expected 
AF wave vectors.

For energy transfers above $E=50$ meV, near the optical phonon cutoff, the background becomes well behaved. It can be described well with a linear radial component and a component in the direction parallel to the detector tubes as described above. The challenge in fitting a radial background is the increased diffusion of signal throughout the Brillouin Zone. The solution takes advantage of the lack of dispersion along the radial direction. A longitudinal and transverse cut through the AF wave vector were fit simultaneously, restricting the background at the AF wave vector to be the same. Essentially, the background is viewed a cone in ${\bf Q}$ with an offset. As such, the longitudinal cut was fit with a Gaussian atop a linear background and the transverse cut was fit with properly constrained Gaussians atop a hyperbola. The number of Gaussians in the transverse cut were chosen empirically, with three Gaussians per side from $50\leq E\leq 100$ meV to accommodate scattering at [$\pm$1,$\pm$1] and two Gaussians per side $E<100$ meV.

%\bibliography{NoEndingPoint}

\begin{thebibliography}{}

\bibitem{PALee} P. A. Lee, N. Nagaosa and X. -G. Wen, Rev. Mod. Phys. {\bf 78}, 17
(2006).

\bibitem{Eschrig}  M. Eschrig, Adv. Phys. {\bf 55}, 47 (2006).

\bibitem{tranquada} J. M. Tranquada, G. Xu, and I. A. Zaliznyak, J. Magn. Magn. Mater.
350, 148 (2014).

\bibitem{kamihara} Y. Kamihara, T. Watanabe, M. Hirano, and H. Hosono,
  J. Am. Chem. Soc. \textbf{130}, 3296 (2008).

\bibitem{Stewart} G. R. Stewart, Rev. Mod. Phys. {\bf 83}, 1589 (2011).

\bibitem{scalapino} D. J. Scalapino, Rev. Mod. Phys. {\bf 84}, 1383 (2012).

\bibitem{dai} P. C. Dai, J. Hu, and E. Dagotto, Nat. Phys. {\bf 8}, 709 (2012).

\bibitem{dai15} P. C. Dai, Rev. Mod. Phys. {\bf 87}, 855(2015).

\bibitem{inosov} D. S. Inosov, C. R. Physique {\bf 17}, 60 (2016).

\bibitem{mdlumsden} M. D. Lumsden, A. D. Christianson, D. Parshall, M. B. Stone, S. E. Nagler, G. J. MacDougall, H. A. Mook, K. Lokshin, T. Egami, D. L. Abernathy, E. A. Goremychkin, R. Osborn, M. A. McGuire, A. S. Sefat, R. Jin, B. C. Sales, and D. Mandrus, Phys. Rev. Lett. {\bf 102}, 107005 (2009).

\bibitem{sxchi} S. Chi, A. Schneidewind, J. Zhao, L. W. Harriger, L. Li, Y. Luo, G. Cao, Z. Xu, M. Loewenhaupt, J. Hu,
and P. C. Dai, Phys. Rev. Lett. {\bf 102}, 107006 (2009).

\bibitem{sli09} S. Li, Y. Chen, S. Chang, J. W. Lynn, L. Li, Y. Luo, G. Cao, Z. Xu, and P. C. Dai, Phys. Rev. B {\bf 79}, 174527 (2009).

\bibitem{dkpratt09} D. K. Pratt, W. Tian, A. Kreyssig, J. L. Zarestky, S. Nandi, N. Ni, S. L. Bud'ko, P. C. Canfield, A. I. Goldman, and R. J. McQueeney, Phys. Rev. Lett.{\bf 103}, 087001 (2009).

\bibitem{adChristianson} A. D. Christianson, M. D. Lumsden, S. E. Nagler, G. J. MacDougall, M. A. McGuire, A. S. Sefat, R. Jin, B. C. Sales, and D. Mandrus, Phys. Rev. Lett. {\bf 103}, 087002 (2009).

\bibitem{dsinosov09} D. S. Inosov, J. T. Park, P. Bourges, D. Sun, Y. Sidis, A. Schneidewind, K. Hradil, D. Haug, C. Lin,
B. Keimer, and V. Hinkov, Nat. Phys. {\bf 6}, 178 (2010).

\bibitem{mywang10} M. Wang, H. Luo, J. Zhao, C. Zhang, M. Wang, K. Marty, S. Chi, J. W. Lynn, A. Schneidewind, S. Li,
and P. C. Dai, Phys. Rev. B {\bf 81}, 174524 (2010).

\bibitem{clester10} C. Lester, J. Chu, J. G. Analytis, T. G. Perring, I. R. Fisher, and S. M. Hayden, Phys. Rev. B {\bf 81}, 064505 (2010).

\bibitem{jtpark10} J. T. Park, D. S. Inosov, A. Yaresko, S. Graser, D. Sun, Ph. Bourges, Y. Sidis, Y. Li, J. -H. Kim, D. Haug, A. Ivanov, K. Hradil, A. Schneidewind, P. Link, E. Faulhaber, I. Glavatskyy, C. Lin, B. Keimer, and V. Hinkov, Phys. Rev. B {\bf 82}, 134503 (2010).

\bibitem{hfli10} H. Li, C. Broholm, D. Vaknin, R. M. Fernandes, D. L. Abernathy, M. B. Stone, D. K. Pratt, W. Tian, Y. Qiu, N. Ni, S. O. Diallo,
J. L. Zarestky, S. L. Bud'ko, P. C. Canfield, and R. J. McQueeney, Phys. Rev. B {\bf 82}, 140503(R) (2010).

\bibitem{dsinosov11} D. S. Inosov, J. T. Park, A. Charnukha, Y. Li, A. V. Boris, B. Keimer, and V. Hinkov
Phys. Rev. B {\bf 83}, 214520 (2011).

\bibitem{mwang11} M. Wang, H. Luo, M. Wang, S. Chi, J. A. Rodriguez-Rivera, D. Singh, S. Chang, J. W. Lynn, and P. C. Dai,
Phys. Rev. B {\bf 83}, 094516 (2011).

\bibitem{lharriger} L. W. Harriger, H. Luo, M. Liu, C. Frost, J. Hu, M. R. Norman, and P. C. Dai, Phys. Rev. B {\bf 84}, 054544 (2011).

\bibitem{kmatan} K. Matan, S. Ibuka, R. Morinaga, S. Chi, J. W. Lynn, A. D. Christianson, M. D. Lumsden, and T. J. Sato, Phys. Rev. B {\bf 82}, 054515 (2010); Phys. Rev. B {\bf 83},  059901 (2011).

\bibitem{msliu12} M. Liu, L. W. Harriger, H. Luo, M. Wang, R. A. Ewings, T. Guidi, H. Park, K. Haule, G. Kotliar, S. M. Hayden, and
P. C. Dai, Nat. Phys. {\bf 8}, 376 (2012).

\bibitem{hqluo12} H. Q. Luo, Z. Yamani, Y. Chen, X. Lu, M. Wang, S. Li, T. A. Maier, S. Danilkin, D. T. Adroja, and P. C. Dai,
 Phys. Rev. B {\bf 86}, 024508 (2012).

\bibitem{gstucker12} G. S. Tucker, R. M. Fernandes, H. Li, V. Thampy, N. Ni, D. L. Abernathy, S. L. Bud'ko, P. C. Canfield, D. Vaknin, J. Schmalian, and R. J. McQueeney, Phys. Rev. B {\bf 86}, 024505 (2012).

\bibitem{mgkim13} M. G. Kim, G. S. Tucker, D. K. Pratt, S. Ran, A. Thaler, A. D. Christianson, K. Marty, S. Calder, A. Podlesnyak, S. L. Bud'ko, P. C. Canfield, A. Kreyssig, A. I. Goldman, and R. J. McQueeney, Phys. Rev. Lett. {\bf 110}, 177002 (2013).

\bibitem{mwang13} M. Wang, C. L. Zhang, X. Lu, G. Tan, H. Luo, Y. Song, M. Wang, X. Zhang, E. A. Goremychkin, T. G. Perring, T. A. Maier, Z. Yin, K. Haule, G. Kotliar, P. C. Dai, Nat. Comms. {\bf 4}, 2874 (2013).

\bibitem{hqluo13} H. Q. Luo, X. Y. Lu, R. Zhang, M. Wang, E. A. Goremychkin, D. T. Adroja, S. Danilkin, G. Deng, Z. Yamani, and P. C. Dai 
 Phys. Rev. B 88, 144516 (2013).   

\bibitem{Ibuka} S. Ibuka, Y. Nambu, T. Yamazaki, M.D. Lumsden, and T. J. Sato, 
Physica C {\bf 507}, 25 (2014). 

\bibitem{MGKim15} M. G. Kim, M. Wang, G. S. Tucker, P. N. Valdivia, D. L. Abernathy, Songxue Chi, A. D. Christianson, A. A. Aczel, T. Hong, T. W. Heitmann, S. Ran, P. C. Canfield, E. D. Bourret-Courchesne, A. Kreyssig, D. H. Lee, A. I. Goldman, R. J. McQueeney, and R. J. Birgeneau, Phys. Rev. B {\bf 92}, 214404 (2015).

\bibitem{cruz} C. de la Cruz, Q. Huang, J. W. Lynn, J. Li, W. Ratcliff II, J. L. Zarestky, H. A. Mook, G. Chen, J. Luo, N. Wang, and P. C. Dai,
Nature (London) {\bf 453}, 899 (2008).

\bibitem{qhunag} Q. Huang, Y. Qiu, W. Bao, M. A. Green, J. W. Lynn, Y. C.
Gasparovic, T. Wu, G. Wu, and X. H. Chen, Phys. Rev. Lett. {\bf 101}, 257003 (2008).

\bibitem{Lester09} C. Lester, J.-H. Chu, J. G. Analytis, S. C. Capelli, A. S.
Erickson, C. L. Condron, M. F. Toney, I. R. Fisher, and S. M. Hayden, Phys. Rev. B {\bf 79}, 144523 (2009).

\bibitem{Pratt09} D. K. Pratt, W. Tian, A. Kreyssig, J. L. Zarestky, S. Nandi,
N. Ni, S. L. Bud'ko, P. C. Canfield, A. I. Goldman, and R. J. McQueeney, Phys. Rev. Lett. {\bf 103}, 087001 (2009).

\bibitem{Christianson09} A. D. Christianson, M. D. Lumsden, S. E. Nagler, G. J.
MacDougall, M. A. McGuire, A. S. Sefat, R. Jin, B. C.
Sales, and D. Mandrus, Phys. Rev. Lett. {\bf 103}, 087002 (2009).

\bibitem{Nandi10} S. Nandi, M. G. Kim, A. Kreyssig, R. M. Fernandes,
D. K. Pratt, A. Thaler, N. Ni, S. L. Bud’ko, P. C. Canfield,
J. Schmalian, R. J. McQueeney, and A. I. Goldman, Phys.
Rev. Lett. 104, 057006 (2010).

\bibitem{hqluo} H. Luo, R. Zhang, M. Laver, Z. Yamani, M. Wang, X. Lu, M. Wang, Y. Chen, S. Li,
S. Chang, J. W. Lynn, P. C. Dai, Phys. Rev. Lett. {\bf 108}, 247002 (2012).

\bibitem{xylu13} X. Lu, H. Gretarsson, R. Zhang, X. Liu, H. Luo, W. Tian, M. Laver, Z. Yamani, Y. -J. Kim, A. H. Nevidomskyy, Q. Si, and P. C. Dai, Phys. Rev. Lett. {\bf 110}, 257001 (2013).

\bibitem{mazin2011n} I. I. Mazin, Nature (London) {\bf 464}, 183-186 (2010).

\bibitem{Hirschfeld} P. J. Hirschfeld, M. M. Korshunov, and I. I. Mazin,
Rep. Prog. Phys. {\bf 74}, 124508 (2011).

\bibitem{Chubukov} A. V. Chubukov, Annu. Rev. Condens. Matter Phys. {\bf 3}, 13 (2012).

\bibitem{khaule08} K. Haule, J. H. Shim, and G. Kotliar, Phys. Rev. Lett. {\bf 100}, 226402 (2008).

\bibitem{qmsi08} Q. Si and E. Abrahams, Phys. Rev. Lett. {\bf 101}, 076401 (2008).

\bibitem{cfang08} C. Fang, H. Yao, W. F. Tsai, J. Hu, and S. A. Kivelson, Phys. Rev. B {\bf 77}, 224509 (2008).

\bibitem{ckxu08} C. Xu, M. M${\rm \ddot{u}}$ller, and S. Sachdev, Phys. Rev. B {\bf 78}, 020501(R) (2008).

\bibitem{CWChu09} C. W. Chu, F. Chen, M. Gooch, A. M. Guloy, B. Lorenz,
B. Lv, K. Sasmal, Z. J. Tang, J. H. Tapp, and Y. Y. Xue,
Physica (Amsterdam) {\bf 469C}, 326 (2009).

\bibitem{Parker10} D. R. Parker, M. J. P. Smith, T. Lancaster, A. J. Steele, I. Franke,
P. J. Baker, F. L. Pratt, M. J. Pitcher, S. J. Blundell, and S. J. Clarke
et al., Phys. Rev. Lett. {\bf 104}, 057007 (2010).

\bibitem{AFWang12} A. F. Wang, X. G. Luo, Y. J. Yan, J. J. Ying, Z. J. Xiang, G. J. Ye,
P. Cheng, Z. Y. Li, W. J. Hu, and X. H. Chen, Phys. Rev. B {\bf 85},
224521 (2012).

\bibitem{GTTan13} G. T. Tan, P. Zheng, X. C. Wang, Y. C. Chen, X. T. Zhang, J. L. Luo, T. Netherton, Y. Song, P. C. Dai, C. L. Zhang, 
and S. L. Li, Phys. Rev. B {\bf 87}, 144512 (2013).   

\bibitem{SLi09a} S. Li, C. de la Cruz, Q. Huang, G. F. Chen, T.-L. Xia, J. L. Luo,
N. L. Wang, and P. C. Dai, Phys. Rev. B {\bf 80}, 020504(R) (2009).

\bibitem{CLZhang14} C. L. Zhang, L. W. Harriger, Z. P. Yin, W. C. Lv, M. Y. Wang, G. T. Tan, Y. Song, D. L. Abernathy, W. Tian, T. Egami, K. Haule, G. Kotliar, and P. C. Dai, Phys. Rev. Lett. {\bf 112}, 217202 (2014).   

\bibitem{Kotliar06} G. Kotliar, S. Y. Savrasov, K. Haule, V. S. Oudovenko, O. Parcollet, and C. A. Marianetti, Rev. Mod. Phys. {\bf 78}, 865
(2006).

\bibitem{Haule10} K. Haule, C.-H. Yee, and K. Kim, Phys. Rev. B {\bf 81}, 195107 (2010).

\bibitem{ZPYin14} Z. P. Yin, K. Haule, G. Kotliar, Nat. Phys. {\bf 10}, 845 (2014).

\bibitem{Spalek} J. Spalek, Acta Physica Polonica A {\bf 111}, 409 (2007).

\bibitem{CLZhang13a} C. L. Zhang, H.-F. Li, Y. Song, Y. Su, G. T. Tan, T. Netherton, C. Redding, S. V. Carr, O. Sobolev, A. Schneidewind, E. Faulhaber, L. W. Harriger, S. L. Li, X. Y. Lu, D.-X. Yao, T. Das, A. V. Balatsky, Th. Br$\rm \ddot{u}$ckel, J. W. Lynn, 
and P. C. Dai, Phys. Rev. B {\bf 88}, 064504 (2013). 

\bibitem{CLZhang13b} C. L. Zhang, R. Yu, Y. Su, Y. Song, M. Y. Wang, G. T. Tan, T. Egami, 
J. A. Fernandez-Baca, E. Faulhaber, Q. Si, and P. C. Dai, 
 Phys. Rev. Lett. {\bf 111}, 207002 (2013). 

\bibitem{CLZhang14a} C. L. Zhang, Y. Song, L.-P. Regnault, Y. Su, M. Enderle, J. Kulda, G. T. Tan, Z. C. Sims, T. Egami, Q. Si, 
and P. C. Dai, Phys. Rev. B  {\bf 90}, 140502(R) (2014). 

\bibitem{ARCS} D. L. Abernathy, M. B. Stone, M. J., Loguillo, M. S. Lucas, O. Delaire, X. L. Tang, T. Y. Lin, and B. Fultz, 
Rev. Sci. Instr. {\bf 83}, 015114 (2012).

\bibitem{SEQUOIA} G. E. Granroth, D. H. Vandergriff, and S. E. Nagler, Physica B {\bf 385-386}, 1104 (2006).

\bibitem{Spyrison12} N. Spyrison, M. A. Tanatar, K. Cho, Y. Song, P. C. Dai, C. L. Zhang, and R. Prozorov, Phys. Rev. B {\bf 86}, 144528 (2012).   

\bibitem{QSWang} Qisi Wang, Yao Shen, Bingying Pan, Xiaowen Zhang, K. Ikeuchi, K. Iida, A. D. Christianson, H. C. Walker, D. T. Adroja, M. Abdel-Hafiez, Xiaojia Chen, D. A. Chareev, A. N. Vasiliev, Jun Zhao, arXiv: 1511.02485. 

\bibitem{richard} P. Richard, T. Sato, K. Nakayama, T. Takahashi, and H. Ding, Rep. Prog. Phys. {\bf 74}, 124512 (2011).

\bibitem{JHZhang} J. H. Zhang, R. Sknepnek, and J. Schmalian, Phys. Rev. B {\bf 82}, 134527 (2010).

\bibitem{Harriger12} L. W. Harriger, M. S. Liu, H. Q. Luo, R. A. Ewings, C. D. Frost, T. G. Perring, and P. C. Dai, Phys. Rev. B {\bf 86}, 140403(R) (2012).

\bibitem{ZPYin11} Z. P. Yin, K. Haule, and G. Kotliar, Nat. Matr. {\bf 10}, 932 (2011).

\bibitem{ZHLiu11} Z.-H. Liu, P. Richard, K. Nakayama, G.-F. Chen, S. Dong, J.-B. He, D.-M. Wang, T.-L. Xia, K. Umezawa,
T. Kawahara, S. Souma, T. Sato, T. Takahashi, T. Qian, Yaobo Huang,
Nan Xu, Yingbo Shi, H. Ding, and S.-C. Wang, Phys. Rev. B {\bf 84}, 064519 (2011).

\bibitem{STCui12} S. T. Cui, S. Y. Zhu, A. F. Wang, S. Kong, S. L. Ju, X. G. Luo, X. H. Chen, G. B. Zhang, 
and Z. Sun, Phys. Rev. B {\bf 86}, 155143 (2012).

\bibitem{QQGe13} Q. Q. Ge, Z. R. Ye, M. Xu, Y. Zhang, J. Jiang, B. P. Xie,
Y. Song, C. L. Zhang, P. C. Dai, and D. L. Feng, Phys. Rev. X {\bf 3}, 011020 (2013).

\bibitem{FCHsu} F.-C. Hsu, J.-Y. Luo, K.-W. Yeh, T.-K. Chen, T.-W. Huang, P. M. Wu, Y.-C. Lee, Y.-L. Huang, Y.-Y. Chu, 
 D.-C. Yan, and M.-K. Wu, Proc. Natl. Acad. Sci. U.S.A. {\bf 105}, 14262 (2008).

\bibitem{CHea} C. He, Y. Zhang, X. F. Wang, J. Jiang, F. Chen, L. X. Yang, Z. R. Ye, Fan Wu, M. Arita, K. Shimada, H. Namatame, M. Taniguchi, X. H. Chen, B. P. Xie, D. L. Feng, J. Phys. Chem. Solids {\bf 72} 479 (2010).

\bibitem{ZRYe} Z. R. Ye, Y. Zhang, F. Chen, M. Xu, J. Jiang, X. H. Niu, C. H. P. Wen, L. Y. Xing, X. C. Wang, C. Q. Jin, B. P. Xie, and D. L. Feng, Phys. Rev. X {\bf 4}, 031041 (2014).

\bibitem{PhonNorm} G. Xu, Z. Xu, and J. M. Tranquada, Rev. Sci. Instr. {\bf 84}, 083906 (2013).

\end{thebibliography}

\end{document}